\documentclass[aps,prd,showpacs, floats, twocolumn,superscriptaddress]{revtex4}
\usepackage{graphicx}  
\usepackage{dcolumn}   
\usepackage{bm}        
\usepackage{amssymb}   
\usepackage{amsmath}   
\usepackage{mathptmx}
\usepackage{xspace}
\usepackage{url}
\usepackage[colorlinks=false,linkcolor=black, citecolor=green, linktoc=page, urlcolor=cyan, 
pdffitwindow=false, draft]{hyperref}
\usepackage{footnote}
\usepackage{float}
\usepackage{bm} 

\def\reff@jnl#1{{\rm#1\/}}
\def\aj{\reff@jnl{AJ}}         
\def\araa{\reff@jnl{ARA\&A}}      
\def\apj{\reff@jnl{ApJ}}        
\def\apjl{\reff@jnl{ApJ}}        
\def\apjs{\reff@jnl{ApJS}}       
\def\aap{\reff@jnl{A\&A}}        
\def\aapr{\reff@jnl{A\&A~Rev.}}     
\def\aaps{\reff@jnl{A\&AS}}       
\def\mnras{\reff@jnl{MNRAS}}      
\def\physrep{\reff@jnl{Physics Reports}}
\def\prd{\reff@jnl{Phys.Rev.D}}     
\def\prl{\reff@jnl{Phys.Rev.Lett}}   
\def\pasp{\reff@jnl{PASP}}       
\def\pasj{\reff@jnl{PASJ}}       
\def\nat{\reff@jnl{Nature}}       
\def\jcap{\reff@jnl{JCAP}}   
\def\memsai{\reff@jnl{MemSAI}} 
\def\na{\reff@jnl{New Astronomy}}       

\usepackage{color}
\def\Sref#1{Sec.~\ref{#1}\xspace}

\def\Fref#1{Fig.~(\ref{#1})\xspace}

\def\Tref#1{Table~\ref{#1}\xspace}

\def\Eref#1{Eq.~(\ref{#1})\xspace}

\def\Aref#1{Appendix~\ref{#1}\xspace}
\def\Cref#1{Chapter~\ref{#1}\xspace}

\def\eg{{e.g.}}
\def\ie{{i.e.}}

\def\anl{Argonne National Laboratory, 9700 South Cass Avenue, Lemont, IL 60439, USA}
\def\upenn{Department of Physics and Astronomy, University of Pennsylvania, Philadelphia, PA 19104, USA}
\def\ethz{Department of Physics, ETH Zurich, Wolfgang-Pauli-Strasse 16, CH-8093 Zurich, Switzerland}
\def\ports{Institute of Cosmology \& Gravitation, University of Portsmouth, Portsmouth, PO1 3FX, UK}
\def\ucl{Department of Physics \& Astronomy, University College London, Gower Street, London, WC1E 6BT, UK}
\def\bnl{Brookhaven National Laboratory, Bldg 510, Upton, NY 11973, USA}
\def\fermilab{Fermi National Accelerator Laboratory, P. O. Box 500, Batavia, IL 60510, USA}
\def\stanford{Department of Physics, Stanford University, 382 Via Pueblo Mall, Stanford, CA 94305, USA}
\def\kipac{Kavli Institute for Particle Astrophysics \& Cosmology, P. O. Box 2450, Stanford University, Stanford, CA 94305, USA}
\def\slac{SLAC National Accelerator Laboratory, Menlo Park, CA 94025, USA}
\def\ifae{Institut de F\'{\i}sica d'Altes Energies, Universitat Aut\`onoma de Barcelona, E-08193 Bellaterra, Barcelona, Spain}
\def\ieec{Institut de Ci\`encies de l'Espai, IEEC-CSIC, Campus UAB, Facultat de Ci\`encies, Torre C5 par-2, 08193 Bellaterra, Barcelona, Spain}
\def\ccap{Center for Cosmology and Astro-Particle Physics, The Ohio State University, Columbus, OH 43210, USA}
\def\ohio{Department of Physics, The Ohio State University, Columbus, OH 43210, USA}
\def\manchester{Jodrell Bank Center for Astrophysics, School of Physics and Astronomy, University of Manchester, Oxford Road, Manchester, M13 9PL, UK}
\def\cambridgekavli{Kavli Institute for Cosmology, University of Cambridge, Madingley Road, Cambridge CB3 0HA, UK}
\def\cambridge{Institute of Astronomy, University of Cambridge, Madingley Road, Cambridge CB3 0HA, UK}
\def\paris{Institut d'Astrophysique de Paris, Univ. Pierre et Marie Curie \& CNRS UMR7095, F-75014 Paris, France}
\def\lina{Laborat\'orio Interinstitucional de e-Astronomia - LIneA, Rua Gal. Jos\'e Cristino 77, Rio de Janeiro, RJ - 20921-400, Brazil}
\def\on{Observat\'orio Nacional, Rua Gal. Jos\'e Cristino 77, Rio de Janeiro, RJ - 20921-400, Brazil}
\def\texas{George P. and Cynthia Woods Mitchell Institute for Fundamental Physics and Astronomy, and Department of Physics and Astronomy,
Texas A\&M University, College Station, TX 77843,  USA}
\def\bosch{Robert Bosch LLC, 4009 Miranda Ave, Suite 225, Palo Alto, CA 94304, USA}
\def\michigan{Department of Physics, University of Michigan, Ann Arbor, MI 48109, USA}
\def\michiganastro{Department of Astronomy, University of Michigan, Ann Arbor, MI 48109, USA}
\def\chicagokavli{Kavli Institute for Cosmological Physics, University of Chicago, Chicago, IL 60637, USA}
\def\maxplanck{Max Planck Institute for Extraterrestrial Physics, Giessenbachstrasse, 85748 Garching, Germany}
\def\munich{University Observatory Munich, Scheinerstrasse 1, 81679 Munich, Germany}
\def\lmu{Department of Physics, Ludwig-Maximilians-Universitaet, Scheinerstr. 1, 81679 Muenchen, Germany}
\def\ctio{Cerro Tololo Inter-American Observatory, National Optical Astronomy Observatory, Casilla 603, La Serena, Chile}
\def\aao{Australian Astronomical Observatory, North Ryde, NSW 2113, Australia}
\def\icra{ICRA, Centro Brasileiro de Pesquisas F\'isicas, Rua Dr. Xavier Sigaud 150, CEP 22290-180, Rio de Janeiro, RJ, Brazil}
\def\jpl{Jet Propulsion Laboratory, California Institute of Technology, 4800 Oak Grove Dr., Pasadena, CA 91109, USA}
\def\ciemat{Centro de Investigaciones Energ\'eticas, Medioambientales y Tecnol\'ogicas (CIEMAT), Madrid, Spain}
\def\uiuc{Department of Physics, University of Illinois, 1110 W. Green St., Urbana, IL 61801, USA}
\def\ncsa{National Center for Supercomputing Applications, 1205 West Clark St., Urbana, IL 61801, USA}
\def\ua{University of Arizona, Department of Physics, 1118 E. Fourth St., Tucson, AZ 85721, USA}
\def\jpl{Jet Propulsion Laboratory, California Institute of Technology, 4800 Oak Grove Dr., Pasadena, CA 91109, USA}
\def\sussex{Department of Physics and Astronomy, Pevensey Building, University of Sussex, Brighton, BN1 9QH, UK}
\def\cluster{Excellence Cluster Universe, Boltzmannstr.\ 2, 85748 Garching, Germany}
\def\barcelona{Instituci\'o Catalana de Recerca i Estudis Avan\c{c}ats, E-08010 Barcelona, Spain}
\def\sepnet{SEPnet, South East Physics Network, (www.sepnet.ac.uk)}

\newcommand{\ngmix}{\texttt{ngmix}}
\def\redmagic{Redmagic}

\begin{document}

\widetext
\leftline{Draft: \today}


\title{Wide-Field Lensing Mass Maps from DES Science Verification Data: \\
Methodology and Detailed Analysis}

\author{V.~Vikram}
\affiliation{\anl}
\affiliation{\upenn}
\author{C.~Chang}
\email{chihway.chang@phys.ethz.ch}
\affiliation{\ethz}
\author{B.~Jain}
\affiliation{\upenn}
\author{D.~Bacon}
\affiliation{\ports}  
\author{A.~Amara}
\affiliation{\ethz}

\author{M.~R.~Becker}
\affiliation{\stanford}
\affiliation{\kipac}    
\author{G.~Bernstein}
\affiliation{\upenn}  
\author{C.~Bonnett}
\affiliation{\ifae}  
\author{S.~Bridle}
\affiliation{\manchester}   
\author{D.~Brout}
\affiliation{\upenn}    
\author{M.~Busha}
\affiliation{\stanford}
\affiliation{\kipac}    
\author{J.~Frieman}
\affiliation{\chicagokavli}
\affiliation{\fermilab}
\author{E.~Gaztanaga}
\affiliation{\ieec}
\author{W.~Hartley}
\affiliation{\ethz} 
\author{M.~Jarvis}
\affiliation{\upenn} 
\author{T.~Kacprzak}
\affiliation{\ethz} 
\author{A.~Kov\'acs}
\affiliation{\ifae} 
\author{O.~Lahav}
\affiliation{\ucl} 
\author{B.~Leistedt}
\affiliation{\ucl} 
\author{H.~Lin}
\affiliation{\fermilab} 
\author{P.~Melchior}
\affiliation{\ccap} 
\affiliation{\ohio} 
\author{H.~Peiris}
\affiliation{\ucl} 
\author{E.~Rozo}
\affiliation{\ua} 
\author{E.~Rykoff}
\affiliation{\kipac} 
\affiliation{\slac} 
\author{C.~S\'anchez}
\affiliation{\ifae} 
\author{E.~Sheldon}
\affiliation{\bnl}   
\author{M.~A.~Troxel}
\affiliation{\manchester} 
\author{R.~Wechsler}
\affiliation{\stanford} 
\affiliation{\kipac} 
\affiliation{\slac}  
\author{J.~Zuntz}
\affiliation{\manchester} 
\author{T.~Abbott}
\affiliation{\ctio} 
\author{F.~B.~Abdalla}
\affiliation{\ucl} 
\author{R.~Armstrong}
\affiliation{\upenn} 
\author{M.~Banerji}
\affiliation{\cambridge} 
\affiliation{\cambridgekavli} 
\author{A.~H.~Bauer}
\affiliation{\ieec} 
\author{A.~Benoit-L{\'e}vy}
\affiliation{\ucl} 
\author{E.~Bertin}
\affiliation{\paris} 
\author{D.~Brooks}
\affiliation{\ucl} 
\author{E.~Buckley-Geer}
\affiliation{\fermilab} 
\author{D.~L.~Burke}
\affiliation{\kipac} 
\affiliation{\slac} 
\author{D.~Capozzi}
\affiliation{\ports} 
\author{A.~Carnero~Rosell}
\affiliation{\lina} 
\affiliation{\on} 
\author{M.~Carrasco~Kind}
\affiliation{\uiuc} 
\affiliation{\ncsa} 
\author{F.~J.~Castander}
\affiliation{\ieec} 
\author{M.~Crocce}
\affiliation{\ieec} 
\author{C.~E.~Cunha}
\affiliation{\bosch} 
\author{C.~B.~D'Andrea}
\affiliation{\ports} 
\author{L.~N.~da Costa}
\affiliation{\lina} 
\affiliation{\on} 
\author{D.~L.~DePoy}
\affiliation{\texas} 
\author{S.~Desai}
\affiliation{\lmu} 
\author{H.~T.~Diehl}
\affiliation{\fermilab} 
\author{J.~P.~Dietrich}
\affiliation{\lmu} 
\affiliation{\cluster}  
\author{J.~Estrada}
\affiliation{\fermilab} 
\author{A.~E.~Evrard}
\affiliation{\michigan} 
\author{A.~Fausti Neto}
\affiliation{\lina} 
\author{E.~Fernandez}
\affiliation{\ifae} 
\author{B.~Flaugher}
\affiliation{\fermilab} 
\author{P.~Fosalba}
\affiliation{\ieec} 
\author{D.~Gerdes}
\affiliation{\michigan} 
\author{D.~Gruen}
\affiliation{\maxplanck} 
\affiliation{\munich} 
\author{R.~A.~Gruendl}
\affiliation{\uiuc} 
\affiliation{\ncsa} 
\author{K.~Honscheid}
\affiliation{\uiuc} 
\affiliation{\ncsa} 
\author{D.~James}
\affiliation{\ctio} 
\author{S.~Kent}
\affiliation{\fermilab} 
\author{K.~Kuehn}
\affiliation{\aao} 
\author{N.~Kuropatkin}
\affiliation{\fermilab} 
\author{T.~S.~Li}
\affiliation{\texas} 
\author{M.~A.~G.~Maia}
\affiliation{\lina} 
\affiliation{\on} 
\author{M.~Makler}
\affiliation{\icra} 
\author{M.~March}
\affiliation{\upenn} 
\author{J.~Marshall}
\affiliation{\texas} 
\author{P.~Martini}
\affiliation{\ccap} 
\affiliation{\ohio} 
\author{K.~W.~Merritt}
\affiliation{\fermilab} 
\author{C.~J.~Miller}
\affiliation{\michigan} 
\affiliation{\michiganastro} 
\author{R.~Miquel}
\affiliation{\ifae} 
\affiliation{\barcelona} 
\author{E.~Neilsen}
\affiliation{\fermilab} 
\author{R.~C.~Nichol}
\affiliation{\ports} 
\author{B.~Nord}
\affiliation{\fermilab} 
\author{R.~Ogando}
\affiliation{\lina} 
\affiliation{\on} 
\author{A.~A.~Plazas}
\affiliation{\bnl} 
\affiliation{\jpl} 
\author{A.~K.~Romer}
\affiliation{\sussex} 
\author{A.~Roodman}
\affiliation{\kipac} 
\affiliation{\slac} 
\author{E.~Sanchez}
\affiliation{\ciemat} 
\author{V.~Scarpine}
\affiliation{\fermilab} 
\author{I.~Sevilla}
\affiliation{\uiuc} 
\affiliation{\ciemat} 
\author{R.~C.~Smith}
\affiliation{\ctio} 
\author{M.~Soares-Santos}
\affiliation{\fermilab} 
\author{F.~Sobreira}
\affiliation{\fermilab} 
\affiliation{\lina} 
\author{E.~Suchyta}
\affiliation{\ccap} 
\affiliation{\ohio} 
\author{M.~E.~C.~Swanson}
\affiliation{\ncsa} 
\author{G.~Tarle}
\affiliation{\lmu} 
\author{J.~Thaler}
\affiliation{\uiuc} 
\author{D.~Thomas}
\affiliation{\ports} 
\affiliation{\sepnet} 
\author{A.~R.~Walker}
\affiliation{\ctio} 
\author{J.~Weller}
\affiliation{\lmu} 
\affiliation{\cluster} 
\affiliation{\maxplanck}  

\date{\today}

\begin{abstract}
Weak gravitational lensing allows one to reconstruct the spatial distribution of the projected mass 
density across the sky. These ``mass maps'' provide a powerful tool for studying cosmology as they
probe both luminous and dark matter. In this paper, we present a weak lensing mass map 
reconstructed from shear measurements in a 139 deg$^2$ area from the Dark Energy Survey 
(DES) Science Verification data. We 
compare the distribution of mass with that of the foreground distribution of galaxies and clusters. 
The overdensities in the reconstructed map correlate well with the distribution of optically
detected clusters. We demonstrate that candidate superclusters and voids along the line of sight 
can be identified, exploiting the tight scatter of the cluster photometric redshifts. 
We cross-correlate the mass map with a foreground magnitude-limited galaxy sample from the 
same data. Our measurement gives results consistent with mock catalogs from N-body simulations 
that include the primary sources of 
statistical uncertainties in the galaxy, lensing, and photo-$z$ catalogs. The statistical significance 
of the cross-correlation is at the 6.8$\sigma$ level with 20 arcminute smoothing. 
We find that the contribution of systematics to the 
lensing mass maps is generally within measurement uncertainties. In this work, we analyze less 
than 3 \% of the final area that will be mapped by the DES; the tools and analysis techniques 
developed in this paper can be applied to forthcoming larger datasets from the survey.
\end{abstract}

\maketitle

\section{Introduction}
\label{sec:intro}
Weak gravitational lensing is a powerful tool for cosmological studies \citep[see][for detailed 
reviews]{bartelmann01, Hoekstra2008}. As light from distant galaxies passes through the mass 
distribution in the Universe, its trajectory gets perturbed, causing the apparent galaxy shapes to 
be distorted. Weak lensing statistically measures this small distortion, or ``shear'', for a large 
number of galaxies to infer the 3D matter distribution. This allows us to constrain cosmological 
parameters and study the distribution of mass in the Universe.

Since its first discovery, the accuracy and statistical precision of weak lensing measurements 
have improved significantly \citep{waerbeke2000, Kaiser2000, Bacon2000, Hoekstra2006, Lin2012,
Heymans2012}. Most of these previous studies constrain cosmology through N-point statistics of 
the shear signal \citep[\eg][]{Bacon2003, Jarvis2006, Semboloni2006, Fu2014, Jee2013, 
2013MNRAS.430.2200K}. In this paper, however, we focus on generating 2D wide-field 
projected mass maps from the measured shear \citep{Waerbeke2014}. These mass maps are 
particularly useful for viewing the non-Gaussian distribution of dark matter in a different way than is 
possible with N-point statistics.

Probing the dark matter distribution in the Universe is particularly important for several reasons. 
Based on the peak statistics from a mass map it is possible to identify dark matter halos and constrain
cosmological parameters \citep[\eg][]{Jain2000, 2008MNRAS.391..435F, 2010MNRAS.402.1049D, 
Kratochvil2010, 2010ApJ...712..992B}. 
Mass maps also allow us to study the connection between baryonic matter (both in stellar and
gaseous forms) and dark matter \citep{Waerbeke2014}. This can be measured by cross correlating light 
maps and gas maps with weak lensing mass maps. Correlation with light maps, which can be constructed 
using observed galaxies, groups and clusters of galaxies etc., can be used to constrain galaxy bias, the 
mass-to-light ratio, and the dependence of these statistics on redshift and environment 
\citep{2012MNRAS.424..553A, 2012MNRAS.426.3369J, 2014MNRAS.442.2534S, 2014ApJ...797..106H}. 
However, one needs to take caution when interpreting the weak lensing mass maps, as the completeness 
and purity of structure detection via these maps is not very high due to their noisy nature 
\citep{2002ApJ...575..640W}.

One other interesting application of the mass map is that it allows us to identify large scale structures 
(both super-clusters and voids) which are otherwise difficult to find \citep[\eg][]{heymans08}. Characterizing 
the statistics of large structures can be a sensitive probe of cosmological models. Structures with masses as 
high or higher than clusters require special attention as the massive end of the halo mass function is very
sensitive to the cosmology \citep{bahcall98, haiman01,holder01}. These rare structures also allows us to 
constrain different theories of gravity \citep{knox06,jain10}. In addition to the study of the largest assemblies 
of mass, the study of number density of the largest voids allows further tests of the $\Lambda$CDM model 
\citep[\eg][]{plionis02}.

Similar mass mapping technique as used in this paper has been previously applied to the Canada-France-Hawaii 
Telescope Lensing Survey (CFHTLenS) as presented in \citet{waerbeke13}. Their work demonstrated the 
potential scientific value of these wide-field lensing mass maps, including measuring high-order moments of the 
maps and cross-correlation with galaxy densities. The total area of the mass map in that work is similar to our 
dataset, though it was divided into four separate smaller fields.

The main goal of this paper is to construct a weak lensing mass map from a contiguous 139 deg$^{2}$ area in the 
Dark Energy Survey\footnote{\url{http://www.darkenergysurvey.org}} \citep[DES,][]{DES2005, 2005IJMPA..20.3121F} 
Science Verification (SV) data, which overlaps with the South Pole Telescope survey (the SPT-E field). The SV data 
were recorded using the newly commissioned wide-field mosaic camera, the Dark Energy Camera 
\citep[DECam;][]{2012PhPro..37.1332D, 2012SPIE.8446E..11F, 2015arXiv150402900F} on the 4m Blanco telescope 
at the Cerro Tololo Inter-American Observatory (CTIO) in Chile. We cross correlate this reconstructed mass map 
with optically identified structures such as galaxies and clusters of galaxies. This work opens up several 
directions for future explorations with these mass maps.

This paper is organized as follows. In \Sref{sec:method} we describe the theoretical foundation and methodology 
for constructing the mass maps and galaxy density maps used in this paper. We then describe in \Sref{sec:data} 
the DES dataset used in this work, together with the simulation used to interpret our results. In \Sref{sec:massmaps} 
we present the reconstructed mass maps. We discuss qualitatively in \Sref{sec:corr-rm-cluster} the correlation of 
these maps with known 
foreground structures found via independent optical techniques. In \Sref{sec:m2l}, we quantify the wide-field 
mass-to-light correlation on different spatial scales using the full field. We show that our results are consistent with
expectations from simulations. In \Sref{sec:sys} we estimate the level of contamination by systematics in our results 
from a wide range of sources. Finally, we conclude in \Sref{sec:conclusion}. For a summary of the main results from 
this work, see the companion paper in PRL \citep{2015arXiv150501871C}. 

\section{Methodology}
\label{sec:method}
In this section we first briefly review the principles of weak lensing in \Sref{sec:weak-basic}. Then, we describe 
the adopted mass reconstruction method in \Sref{sec:ks}. Finally in \Sref{sec:fgmaps}, we describe our method of 
generating galaxy density maps. The galaxy density maps are used as independent mass tracers in this work to 
help confirm the signal measured in the weak lensing mass maps. 

\subsection{Weak gravitational lensing}
\label{sec:weak-basic}
When light from galaxies passes through a foreground mass distribution, the resulting bending of light leads to the 
galaxy images being distorted \citep[\eg][]{bartelmann01}. This phenomenon is called gravitational lensing. The 
local mapping between the source ($\bm{\beta}$) and image ($\bm{\theta}$) plane coordinates (aside from an overall 
displacement) can be described by the lens equation:
\begin{equation}
\bm{\beta} - \bm{\beta_{0}}  = A(\bm{\theta}) (\bm{\theta}-\bm{\theta_{0}}),
\end{equation}
where $\bm{\beta_{0}}$ and $\bm{\theta_{0}}$ is the reference point in the source and the image plane. $A$ is the 
Jacobian of this mapping, given by
\begin{equation}
A(\bm{\theta}) = (1-\kappa) \begin{pmatrix}
1 - g_1 & - g_2 \\
-g_2 & 1+g_1
\end{pmatrix}, 
\label{eq:jacobian1}
\end{equation}
where $\kappa$ is the convergence, $g_i = \gamma_i / (1-\kappa)$ is the reduced shear and $\gamma_i$ is the 
shear. $i=1,2$ refers to the 2D coordinates in the plane. The factor $(1-\kappa)$ causes galaxy images 
to be dilated or reduced in size, while the terms in the matrix cause distortion in the image shapes.
Under the Born approximation, which assumes that the deflection of the light rays due to the lensing effect is small, 
$A$ is given by \citep[\eg][]{bartelmann01}
\begin{equation}
A_{ij}(\bm{\theta}, r) = \delta_{ij} - \psi_{,ij},
\label{eq:jacobian2}
\end{equation}
where $\psi$ is the lensing deflection potential, or a weighted projection of the gravitational potential along the line of sight. 
For a spatially flat Universe, it is given by the line of sight integral of the 3D gravitational potential $\Phi$ \citep{2015PhR...568....1J},
\begin{equation}
\psi\left(\bm{\theta}, r\right) = 2 \int_0^r{\mathrm{d}r^\prime \frac{r - r^\prime}{rr^{\prime}} \Phi\left(\bm{\theta}, r^\prime\right)},
\label{eq:2dpotential}
\end{equation}
where $r$ is the comoving distance.
Comparison of \Eref{eq:jacobian2} with \Eref{eq:jacobian1} gives
\begin{equation}
\kappa = \frac{1}{2}\nabla^2 \psi;
\label{eq:kappaphi}
\end{equation}
\begin{equation}
\bm{\gamma} =\gamma_1 + i \gamma_2 = \frac{1}{2}\left(\psi_{,11} - \psi_{,22}\right) + i \psi_{,12}.
\label{eq:shear}
\end{equation}
For the purpose of this paper, we use the Limber approximation which lets us use the Poisson equation 
for the density fluctuation $\delta = (\Delta - \bar{\Delta})/\bar{\Delta}$ (where $\Delta$ and $\bar{\Delta}$ 
are the 3D density and mean density respectively):
\begin{equation}
\nabla^2\Phi = \frac{3H_0^2\Omega_m}{2a} \delta,
\end{equation}
where $a$ is the cosmological scale factor. \Eref{eq:2dpotential} and \Eref{eq:kappaphi} give the 
convergence measured at a sky coordinate 
$\theta$ from sources at comoving distance $r$: 
\begin{equation}
\kappa(\bm{\theta},r) = \frac{3H_0^2\Omega_m}{2} \int_0^r{\mathrm{d}r^\prime \frac{r^\prime (r-r^\prime) }{r} 
\frac{\delta\left(\bm{\theta}, r^\prime\right)} {a(r^\prime)}}.
\label{eq:kappa}
\end{equation}
We can generalize to sources with a distribution in comoving distance (or redshift) $f(r)$ as: 
$\kappa(\bm{\theta}) = \int{\kappa(\bm{\theta}, r) f(r) \mathrm{d}r}$.
That is, a $\kappa$ map constructed over a region on the sky gives us the integrated mass density fluctuation 
in the foreground of the $\kappa$ map weighted by the lensing weight $p(r^\prime)$, which is itself an integral over $f(r)$: 
\begin{equation}
\kappa(\bm{\theta}) = \frac{3H_0^2\Omega_m}{2} \int_{0}^{r} \mathrm{d}r^\prime p(r^\prime) r^\prime \frac{\delta\left(\bm{\theta}, r^\prime\right)}{a(r^\prime)},
\label{eq:kappa2}
\end{equation}
with
\begin{equation} 
p(r^\prime) = \int_{r^\prime}^{r_{H}}{\mathrm{d}r f(r) \frac{r - r^\prime}{r}},
\label{eq:lens_weight}
\end{equation}
where $r_{H}$ is the comoving distance to the horizon. For a specified cosmological model and $f(r)$ specified 
by the redshift distribution of source galaxies, the above equations provide the basis for predicting the statistical 
properties of $\kappa$. 

\subsection{Mass maps from Kaiser-Squires reconstruction}
\label{sec:ks}
In this paper we perform weak lensing mass reconstruction based on the method developed in 
\citet{ks93}. The Kaiser-Squires (KS) method is known to work well up to a constant additive factor as 
long as the structures are in the linear regime \citep{waerbeke13}. 
In the non-linear regime (scales corresponding to clusters or smaller structures) improved 
methods have been developed to recover the mass distribution \citep[\eg][]{bartelmann96,Bridle1998}. In this 
paper we are interested in the mass distribution on large scales; we can therefore restrict 
ourselves to the KS method. The KS method works as follows.  
The Fourier transform of the observed shear, $\bm{\hat{\gamma}}$, relates to the Fourier transform of 
the convergence, $\hat{\kappa}$ through
\begin{equation}
\hat{\kappa}_{\bm{\ell}} = D^*_{\bm{\ell}} \hat{\bm{\gamma}}_{\bm{\ell}},
\label{eq:ks_ft}
\end{equation}
\begin{equation}
D_{\bm{\ell}} = \frac{\ell_1^2 - \ell_2^2 + 2 i \ell_1 \ell_2}{|{\bm{\ell}}|^2},
\label{eq:D}
\end{equation}
where $\ell_i$ are the Fourier counterparts for the angular coordinates $\theta_i$, $i=1, 2$ represent the 
two dimensions of sky coordinate. The above equations hold true for $\bm{\ell}\neq 0$. In practice we apply a 
sinusoidal projection of sky with a reference point at RA=71.0 deg and then pixelize the 
observed shears with a pixel size of 5 arcmin before Fourier transforming. Given that we mainly focus 
on scales less than a degree in this paper, the errors due to the projection 
is small \citep{waerbeke13}.

The inverse Fourier transform of \Eref{eq:ks_ft} gives the convergence for the observed field 
in real space. Ideally, the imaginary part of the inverse Fourier transform will be zero as the 
convergence is a real quantity. However, noise, systematics and masking causes the reconstruction to be 
imperfect, with non-zero imaginary convergence as we will quantify in \Sref{sec:sim_test}. The real and 
imaginary parts of the reconstructed convergence are referred to as the E- and B-mode of $\kappa$, respectively.
In our reconstruction procedure we set shears to zero in the masked regions \citep{2015PhRvD..91f3507L}. We 
later quantify the effect of this step in \Sref{sec:sim_test}. 

One of the issues with the KS inversion is that the uncertainty in the reconstructed convergence is 
formally infinite for a discrete set of noisy shear estimates. This is because the statistically uncorrelated 
ellipticities of galaxies result in a white noise power spectrum which integrates to infinity for large spatial 
frequencies. Therefore we need to remove the high frequency components. For a 
Gaussian filter of size $\sigma$ the covariance of the statistical noise in the convergence map can be 
written as \citep{waerbeke00} 
\begin{equation}
\langle \kappa(\bm{\theta}) \kappa(\bm{\theta^\prime}) \rangle = \frac{\sigma_\epsilon^2}{4 \pi \sigma^2 n_g} 
\exp\left( - \frac{|\bm{\theta} - \bm{\theta^\prime}|^2}{2 \sigma^2}\right),
\label{Eq:error}
\end{equation}
where $\sigma_{\epsilon}$ is the standard deviation of the single component ellipticity (which contains the intrinsic shape 
noise and measurement noise) and $n_g$ is the number density of the source galaxies.
\Eref{Eq:error} implies that the shape noise contribution to the convergence map reduces with 
increasing size of the Gaussian window and number density of the background source galaxies.

\subsection{Lensing-weighted galaxy density maps}
\label{sec:fgmaps}

In addition to the mass map generated from weak lensing measurements in \Sref{sec:ks}, we also generate 
mass maps based on the assumption that galaxies are linearly biased tracers of mass in the foreground. In 
particular, we study two galaxy samples: the general field galaxies and the Luminous Red Galaxies (LRGs). 
Properties of the samples used in this work such as the redshift distribution, magnitude distribution etc. are 
described in \Sref{sec:sample}. To compare with the weak lensing mass map, we assume that the bias is
constant. However, bias may change with spatial scale, redshift, magnitude and other galaxy properties. 
This can introduce differences between the weak lensing mass map and foreground map. In this paper we
neglect such effects since we mostly focus on large scales ($\gtrsim 5-10$ arcmin at $z\sim0.35$) 
where the departures from linear bias are small \citep{2007A&A...461..861S}.

Based on a given sample of mass tracer we generate a weighted foreground map ($\kappa_g$) after applying 
an appropriate lensing weight to each galaxy before pixelation. In principle the weight increases the signal-to-noise 
(S/N) of the cross-correlation between the lensing mass map and the foreground density map. The lensing
weight (\Eref{eq:lens_weight}) depends on the comoving distance to the source and lens, and the distance 
between them. To generate the weighted galaxy density map, we first generate a 3D
grid of the galaxies. We estimate the density contrast in each of these cells as follows:
\begin{equation}
\delta_g^{ijk} = \frac{n_{ijk} - \bar{n}_{k}}{\bar{n}_{k}}
\label{eq:unweighted_fg}
\end{equation}
where $(i,j)$ is the pixel index in the projected 2D sky and $k$ is the pixel index in the redshift direction. 
$n_{ijk}$ is the number of galaxies in the $ijk^{th}$ cell and $\bar{n}_{k}$ is the average number of 
galaxies per pixel in the $k^{th}$ redshift bin. This 3D grid of galaxy density fluctuations 
will be used to estimate $\kappa_g$ according to the discrete version of \Eref{eq:kappa2},
\begin{equation}
\kappa_g^{ij} = \frac{3H_0^2\Omega_m}{2c^2} \sum_k{\Delta_{z} \frac{\delta_k^{3D} d_k}{a_k} \sum_{l>k}{\frac{(d_l-d_k) f_l}{d_l}} },
\label{eq:kg}
\end{equation}
where $\kappa_g^{ij}$ is the weighted foreground map at the pixel $(i,j)$; $k$ and $l$ represent indices 
along the redshift direction for lens and source, $\Delta_{z}$ is the physical size of the redshift bin, $d_l$ 
is the angular diameter distance to source, $f_l$ is the probability density of the source redshift distribution 
at redshift $l$ and $\delta_k^{3D}$ is the foreground density fluctuation at angular diameter distance 
$d_k$. In this work, use a single source redshift bin and $\Delta_{z}=0.1$ for the lens sample. We adopt the 
following cosmological parameters:
$\Omega_{m}=0.3$, $\Omega_{\Lambda}=0.7$, $\Omega_{k}=0.0$, $h=0.72$. 
Our results depend very weakly on the exact values of these cosmological parameters. 

\section{Data and simulations}
\label{sec:data}

The measurements in this paper are based on 139 deg$^{2}$ of data in the SPT-E field, observed as part 
of the Science Verification (SV) data from DES. The SV data were taken during the period of November
2012 -- February 2013 before the official start of the science survey. The data were taken shortly after 
DECam commissioning and were used to test survey operations and assess data quality. Five optical
filters ($grizY$) were used throughout the survey, with typical exposure times being 90 sec for $griz$ 
and 45 sec for $Y$. The final median depth estimates of this data set in our region of interest are
$g\sim24.0$, $r\sim23.9$, $i\sim23.0$ and $z\sim22.3$ (10-$\sigma$ galaxy limiting magnitude).

Below we introduce in \Sref{sec:data_sva1} the relevant data used in this work. Then we define in 
\Sref{sec:sample} two subsamples of the SV data that we identify as ``foreground (lens)'' and 
``background (source)'' galaxies for the main analysis of the paper.
In \Sref{sec:bcc-data} we introduce the simulations we use to interpret our measurements. 

\subsection{The DES SVA1 Gold galaxy catalogs}
\label{sec:data_sva1}

All galaxies used for foreground catalogs and lensing measurements are drawn from the DES SVA1 
Gold Catalog (Rykoff et al., in preparation) and several extensions to it. The main catalog is a product 
of the DES Data Management (DESDM) pipeline version ``SVA1'' (Yanny et al., in preparation). The 
DESDM pipeline, as described in \citet{2006SPIE.6270E..23N, 2011arXiv1109.6741S, 2012ApJ...757...83D, 
2012SPIE.8451E..0DM}, begins with initial image processing on single-exposure images and astrometry 
measurements from the software package \textsc{SCAMP} \citep{2006ASPC..351..112B}. The
single-exposure images were then stacked to produce co-add images using the software package 
\textsc{SWARP} \citep{2002ASPC..281..228B}. Basic object detection, point-spread-function (PSF) 
modelling, star-galaxy classification \footnote{We adopt the ``{\tt MODEST\_CLASS}'' 
classifier, which is a new classifier used for SVA1 Gold that has been developed empirically and 
tested on DES imaging of COSMOS fields with Hubble Space Telescope ACS imaging.} and photometry
were done on the individual images as well as the co-add images using software packages 
\textsc{SExtractor} \citep{1996A&AS..117..393B} and \textsc{PSFEx} \citep{2011ASPC..442..435B}. 
The full SVA1 Gold dataset consists of 254.4 deg$^2$ with $griz$-band coverage, and $223.6$
deg$^2$ for $Y$ band. The main science goal for this work is to reconstruct wide-field mass maps; as 
a result, we use the largest continuous region in the SV data: 139 deg$^{2}$ in the SPT-E field.

The SVA1 Gold Catalog is augmented by: a photometric redshift catalog, two galaxy shape catalogs, 
and an LRG catalog. These catalogs are described below.

\subsubsection{Photometric redshift catalog}
\label{sec:photoZ}
In this work we use the photometric redshift (photo-$z$) estimated with the Bayesian Photometric 
Redshifts (BPZ) code \citep{2000ApJ...536..571B,2006AJ....132..926C}. The photo-$z$'s are
used to select the main foreground and background sample (see \Sref{sec:sample}). The details 
and capabilities of BPZ on early DES data were already presented in \citet{2014MNRAS.445.1482S}, 
where it showed good performance among template-based codes. The primary set of
templates used contains the \citet{1980ApJS...43..393C} templates, two starburst templates from 
\citet{1996ApJ...467...38K} and two younger starburst simple stellar population templates from
\citet{2003MNRAS.344.1000B}, added to BPZ in \citet{2006AJ....132..926C}. We calibrate the 
Bayesian prior by fitting the empirical function $\Pi(z,t|m_0)$ proposed in \citet{2000ApJ...536..571B}, 
using a spectroscopic sample matched to DES galaxies and weighted to mimic the photometric 
properties of the DES SV sample used in this work. As tested in \citet{2014MNRAS.445.1482S}, 
the bias in the photo-$z$ estimate is $\sim$0.02, with 68\% scatter $\sigma_{68}\sim0.1$ and 
the 3$\sigma$ outlier fraction $\sim$2\%. For this work, we use $z_{mean}$, the mean of the 
Probability Distribution Function (PDF) output from BPZ as a single-point estimate of the 
photo-$z$ to separate our galaxies into the foreground and background samples. Other photo-$z$
codes used in DES have been run on the same data. For this work we have also checked our main 
results in \Sref{sec:m2l} using an independent Neural Network code \citep[Skynet;][]{2013arXiv1312.1287B, 
2013ascl.soft12007G}. We found that BPZ and Skynet gives consistent results (within 1$\sigma$) in 
terms of the cross-correlation between the weak lensing mass maps and the foreground galaxy map.

\subsubsection{Shape catalogs}

Based on the SVA1 data, two shear catalogs were produced and tested extensively in Jarvis et al. (in preparation): 
the \texttt{ngmix} \footnote{The open source code can be downloaded at: \url{https://github.com/esheldon/ngmix}} 
(version 011) catalog and the \texttt{im3shape} \footnote{The open source code can be downloaded 
at: \url{https://bitbucket.org/joezuntz/im3shape/}} (version 9) catalog. The main results shown in our paper are 
based on the \texttt{ngmix} catalog, but we also cross-check with the \texttt{im3shape} catalog to confirm that the 
results are statistically consistent. These catalogs are slightly earlier version from that described in Jarvis et al. 
(in preparation). 

The PSF model for both methods are based on the single-exposure PSF models from \textsc{PSFEx}. 
\textsc{PSFEx} models the PSF as a linear combination of small images sampled on an ad hoc pixel grid, with
coefficients that are the terms of a second-order polynomial of pixel coordinates in each DECam CCD.

\ngmix\ \citep{2014MNRAS.444L..25S} is a general tool for fitting morphological models to images of 
astronomical objects. For the 
galaxy model, \ngmix\ supports various options including exponential disk and de Vaucouleurs' profile 
\citep{1948AnAp...11..247D}, all of which are implemented approximately as a sum of Gaussians
\citep{2013PASP..125..719H}. Additionally, any number of Gaussians can be fit. These Gaussian fits 
can either be completely free or constrained to be co-centric and co-elliptical. For the DES SV galaxy
images, we used the exponential disk model. For the PSF fitting, an Expectation Maximization 
\citep{Dempster77maximumlikelihood} approach is used to model the PSF as a sum of three free 
Gaussians. Shear estimation was carried out using by jointly fitting multiple images in $r, i, z$ bands. The 
multi-band approach enabled a larger effective galaxy number density compared to the \texttt{im3shape} 
catalog, which is based on single-band images in the current version.
 
The \texttt{im3shape} \citep{2013MNRAS.434.1604Z} implementation in this work estimates shapes by jointly 
fitting a parameterized galaxy model to all of the different single-exposure $r$-band images, finding the maximum 
likelihood solution. Calibration for bias in the shear measurement associated with noise \citep{2012MNRAS.425.1951R, 
2012MNRAS.427.2711K} is applied. An earlier version of this code (run on the co-add images instead of 
single-exposures) has been run on the SV cluster fields for cluster lensing studies \citep{2015MNRAS.449.2219M}.

Details for both shape catalogs and the tests performed on these catalogs can be found in Jarvis et al. (in preparation). 
Both shear catalogs have been tested and shown to pass the requirements for SV cosmic shear measurement, which 
is much more stringent than what is required in this paper.  As our analysis is insensitive to the overall multiplicative 
bias in the shear measurements, we adopt the ``conservative additive'' selection; this results 
in small additive systematic uncertainties, but possibly some moderate multiplicative systematic uncertainties. 
For \texttt{ngmix}, this selection removes galaxies with S/N$<$20 and very small galaxies (Gaussian sigma smaller 
than the pixel scale). For \texttt{im3shape}, it removes galaxies with S/N$<$15. In both cases, there were many 
other selections applied to both catalogs to remove stars, spurious detections, poor measurements, and various 
other effects that significantly biased shear estimates for both catalogs. 

\subsubsection{The red-sequence Matched filter Galaxy Catalog (Redmagic)}
\label{sec:lrg}

We use the DES SV red-sequence Matched-filter Galaxy Catalog (\redmagic\, Rozo et al., in preparation) 
v6.3.3 in this paper as one of the foreground samples. The objects in this catalog are photometrically 
selected luminous red galaxies (LRGs). We use the terms \redmagic\ galaxies and LRG interchangeably. 
Specifically, \redmagic\ uses the Redmapper-calibrated model for the color of red-sequence galaxies 
as a function of magnitude and redshift \citep{rykoff14}.  This model is used to find the best-fit photometric 
redshift for all galaxies irrespective of type, and the $\chi^2$ goodness-of-fit of the model is computed. 
For each redshift slice, all galaxies fainter than some minimum luminosity threshold $L_{\rm min}$ 
are rejected. In addition, \redmagic\ applies a $\chi^2$ selection $\chi^2 \leq \chi_{\rm max}^2$, where the 
$\chi_{\rm max}^2$ as a function of redshift is chosen to ensure that the resulting galaxy sample has 
a nearly constant space density $\bar{n}$. In this work, we set $\bar{n}=10^{-3} h^3 \rm{Mpc}^{-3}$. 
We assume flat $\Lambda$ CDM model with cosmological parameters $\Omega_{\Lambda}=0.7$, 
$h=100$ (varying these parameters does not change the results significantly). The luminosity selection is 
$L\geq0.5L_*(z)$, where the value of $L_*(z)$ at $z$=0.1 is set to match the Redmapper definition 
for SDSS, and the redshift evolution for $L_*(z)$ is that predicted using a simple passive evolution 
starburst model at $z=3$ \citep{2003MNRAS.344.1000B}. We use the \redmagic\ sample because of the 
exquisite photometric redshifts 
of the \redmagic\ galaxy catalog: \redmagic\ photometric redshifts are nearly unbiased, with a scatter
$\sigma_z/(1+z)\approx 1.7\%$, and a $\approx 1.7\%$ $4\sigma$ redshift outlier rate. We refer the 
reader to Rozo et al. (in preparation) for further details of this catalog.

\subsection{Foreground and background galaxy samples selection}
\label{sec:sample}

As described in \Sref{sec:intro}, the main goal of this paper is to construct a projected mass map at 
a given redshift via weak lensing and to show that the mass map corresponds to real structures, or 
mass, in the foreground line-of-sight. For that purpose, we define two galaxy samples in this study 
--- the background ``source'' sample which is lensed by foreground mass, and the foreground ``lens'' 
sample that traces the foreground mass responsible for the lensing. We wish to construct a weak lensing 
mass map from the background sample according to \Sref{sec:ks} and compare it with the mass map 
generated from the foreground galaxy density map according to \Sref{sec:fgmaps}.

We choose to have the two samples separated at redshift $\sim0.55$ in order to have a sufficient number of galaxies in 
both samples. Given that the photo-$z$ training sample of our photo-$z$ catalog does not extend to the same redshift 
and magnitude range as our data, we exclude objects with photo-$z$ outside the range 0.1--1.2. The final foreground 
and background sample are separated by the photo-$z$ selection of $0.1<z<0.5$ and $0.6<z<1.2$. Note that the \redmagic\ 
foreground galaxy sample has an additional redshift threshold $z>0.2$.

The main quantity of interest for the background galaxy sample is the shear measured for each galaxy. Since 
the background sample only serves as a ``backlight'' for the foreground structure we are interested in, it need not be 
complete. Therefore the most important selection criteria for the background sample is to use galaxies with accurate 
shear measurements. Our source selection criteria are based on extensive tests of shear catalog as described in 
Jarvis et al. (in preparation). After applying the conservative selection of background galaxies and our background redshift 
selection we are left with 1,111,487 galaxies (2.22/arcmin$^2$) for \texttt{ngmix} and 1,013,317 galaxies (2.03/arcmin$^2$) 
for \texttt{im3shape}. 

\begin{figure}[H]
  \begin{center}
  \includegraphics[scale=0.5]{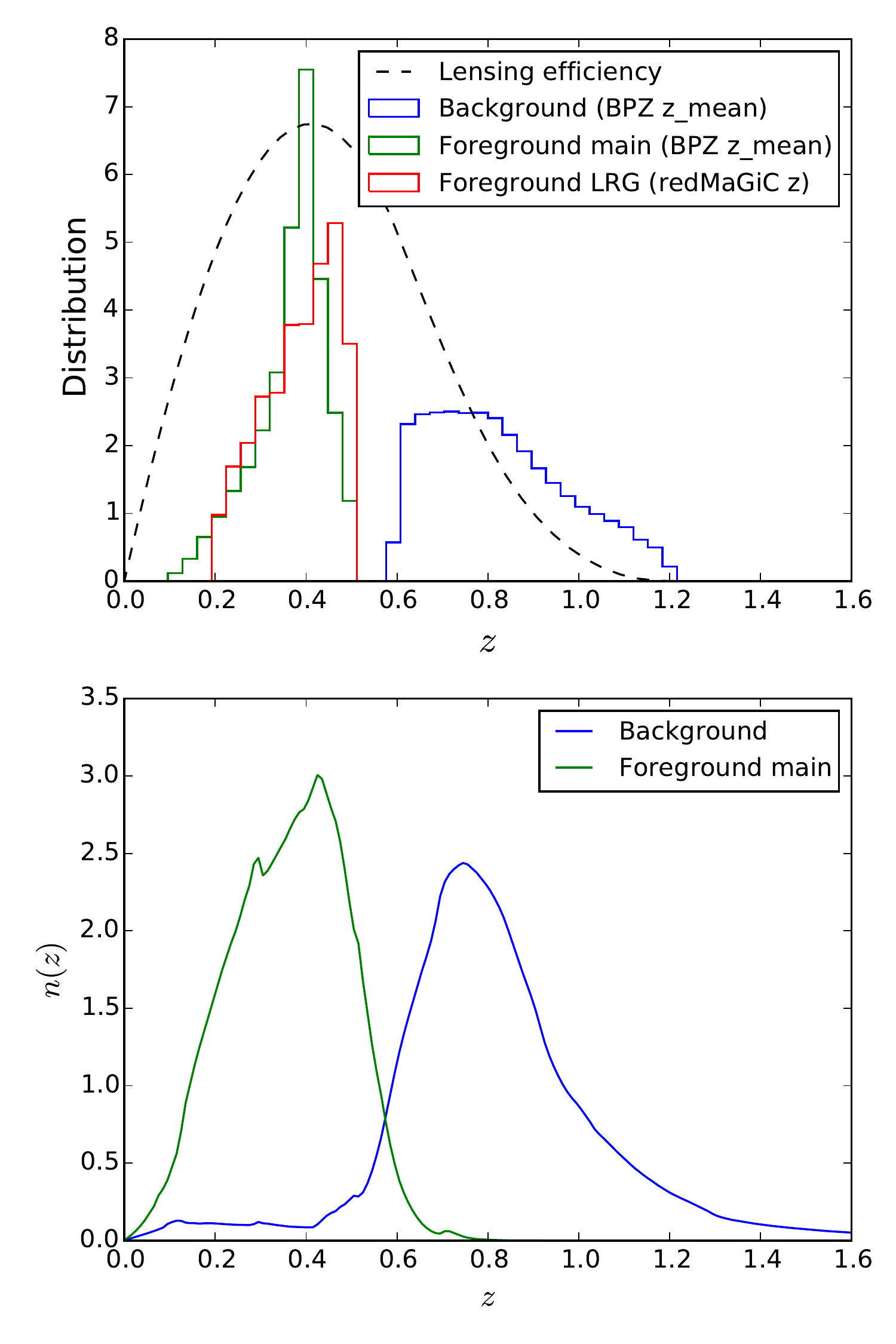}
      \vspace{-0.3in}  
  \end{center}
  \caption{Distributions of the single-point photo-$z$ estimates for the background and foreground samples used in this paper are 
  shown in the top panel, overlaid by the lensing efficiency (\Eref{eq:lens_weight}) corresponding to the background sample. The 
  background and the foreground main sample uses the mean of the PDF from BPZ for single-point 
  estimates, while the LRG redshift estimate comes independently from \redmagic\ (see \Sref{sec:lrg}). The bottom panel 
  shows the corresponding $n(z)$ of the background and foreground main sample given by BPZ. These come from the sum of 
  the PDF for all galaxies in the samples.}
\label{fig:photoZ_dist_sample_ng}
\end{figure}

\begin{table*}
\begin{center}
\caption{Catalogs and selection criteron used to construct the foreground and background sample for this work, and the number of 
galaxies in each sample after all the selections are applied. All catalogs are based on the DES SVA1 dataset. We use the Source 
Extractor MAG\_AUTO parameter for the $i$-band magnitude.}
\begin{tabular}{lcccc}
\\
\hline 
& \multicolumn{2}{c}{Background} & Foreground main & Foreground LRG \\ \hline
Input catalog & \texttt{ngmix011}   & \texttt{im3shape}    & SVA1 Gold &  \redmagic\ \\ \hline   
Photometric redshift & \multicolumn{2}{c}{0.6$<$z$<$1.2} & 0.1$<$z$<$0.5 & 0.2$<$z$<$0.5 \\
Others &  \multicolumn{2}{c}{``conservative additive'' }& $i<$22 & constant density   \\ 
             &                    & &   & $10^{-3}$ $(h^{-1} Mpc)^{-3}$ \\ \hline
Number of galaxies &  1,111,487 & 1,013,317 &  1,106,189 & 28,033  \\   
Number density (arcmin$^{-2}$)& 2.22  & 2.03&  2.21  & 0.056  \\  
Mean redshift   &  0.826 & 0.825  & 0.367  &  0.385  \\  
\hline
\end{tabular}
\label{tab:sample_selection}
\end{center}
\end{table*}

The foreground sample in this work serves as the tracer of mass. Thus it is important to construct a magnitude-limited 
sample for which the number density is affected as little as possible by external factors. The main physical factors that 
contribute to variation in the galaxy number density are the spatial variation in depth and seeing. Both effects can 
introduce spatial variation in the foreground galaxy number density, which can be wrongly identified as foreground 
mass fluctuations. We test both effects in \Aref{sec:foreground_select}. Two subsamples are used in this work as 
foreground samples: the ``main'' foreground sample and the LRG foreground sample. While the space density of 
LRGs is significantly lower than that of the main sample, they are better tracers of galaxy clusters and groups, so 
we use them to check our results. The main foreground sample includes all the galaxies with $i <22$ and the LRG 
sample includes the LRGs in the \redmagic\ LRG catalog with $i < 22$. This magnitude selection is based on tests 
described in \Aref{sec:depth} to ensure that our sample is shallower than the limiting magnitude for all regions of sky 
under study. The final main foreground sample contains 1,106,189 galaxies (2.21/arcmin$^{2}$), while the LRG sample 
contains 28,033 galaxies (0.05/arcmin$^{2}$). \Tref{tab:sample_selection} summarizes all the selection criteria applied 
on the three main samples used in this work.

\Fref{fig:photoZ_dist_sample_ng} shows the distributions of the single-point photo-$z$ estimates ($z_{mean}$) for 
the final foreground and background samples overlaid by the lensing efficiency corresponding to the background 
sample (top panel), and the $n(z)$ (from the BPZ code) for the background and main 
foreground sample (bottom panel). 
Note that the background galaxy number density is much lower than the number density of all galaxies in the 
\texttt{ngmix011} and \texttt{im3shape} catalogs, as we have made stringent redshift selections to avoid overlap 
between the foreground and background samples. 

\subsection{Mock catalogs from simulations}
\label{sec:bcc-data}

We use the simulations primarily as a tool to understand the impact of various effects on the 
expected signal, and a sanity check to confirm that our measurement method is producing reasonable results.
We use a set of simulated galaxy catalogs ``Aardvark v1.0c'' 
developed for the DES collaboration \citep{2013AAS...22134107B}. The full catalog covers
approximately 1/4 of the sky and is 
complete to the final expected DES depth.
 
The heart of the galaxy catalog generation is the algorithm Adding Density Determined Galaxies to Lightcone 
Simulations \citep[ADDGALS;][]{2013AAS...22134107B}, which aims at generating a galaxy catalog that matches 
the luminosities, colors, and clustering properties of the observed data. The simulated galaxy catalog is based
on three flat $\Lambda$CDM dark matter-only N-body simulations, one each of a 1050 Mpc/h, 2600 Mpc/h and 
4000 Mpc/h boxes with $1400^3$, $2048^3$ and $2048^3$ particles respectively. These boxes were run
with \verb+LGadget-2+ \citep{2005MNRAS.364.1105S} with \verb+2LPTic+ initial conditions from 
\citep{2006MNRAS.373..369C} and \verb+CAMB+ \citep{2002PhRvD..66j3511L}. From an input luminosity 
function, galaxies are drawn and then assigned to a position in the dark matter simulation volume according 
to a statistical prescription of the relation between the galaxy's magnitude, redshift and local dark matter density. 
The prescription is derived from a high-resolution simulation using SubHalo Abundance Matching techniques
\citep{2006ApJ...647..201C, 2013ApJ...771...30R, 2013AAS...22134107B}. Next, photometric properties are 
assigned to each galaxy, where the magnitude-color-redshift distribution is designed to reproduce the observed 
distribution of SDSS DR8 \citep{2011ApJS..193...29A} and DEEP2 \citep{2013ApJS..208....5N} data. The size 
distribution of the galaxies is magnitude-dependent and
modelled from a set of deep ($i\sim$26) SuprimeCam $i$-band images, which were taken at with seeing 
conditions of 0.6" \citep{2014MNRAS.440.2191S}. Finally, the weak lensing parameters ($\kappa$ and $\bm{\gamma}$) in 
the simulations are
based on the ray-tracing algorithm Curved-sky grAvitational Lensing for Cosmological Light conE simulatioNS
\citep[CALCLENS;][]{2013MNRAS.435..115B}. The ray-tracing resolution is accurate to $\simeq 6.4$ arcseconds, 
sufficient for the usage in this work.

Aside from the intrinsic uncertainties in the modelling in the mock galaxy catalog (related to the input parameters 
and uncertainty in the galaxy-halo connection), there are also many real-world effects that are not included 
in these simulations, including as depth variation, seeing variation and shear measurement uncertainties. 

\begin{figure*}
\includegraphics[scale=1]{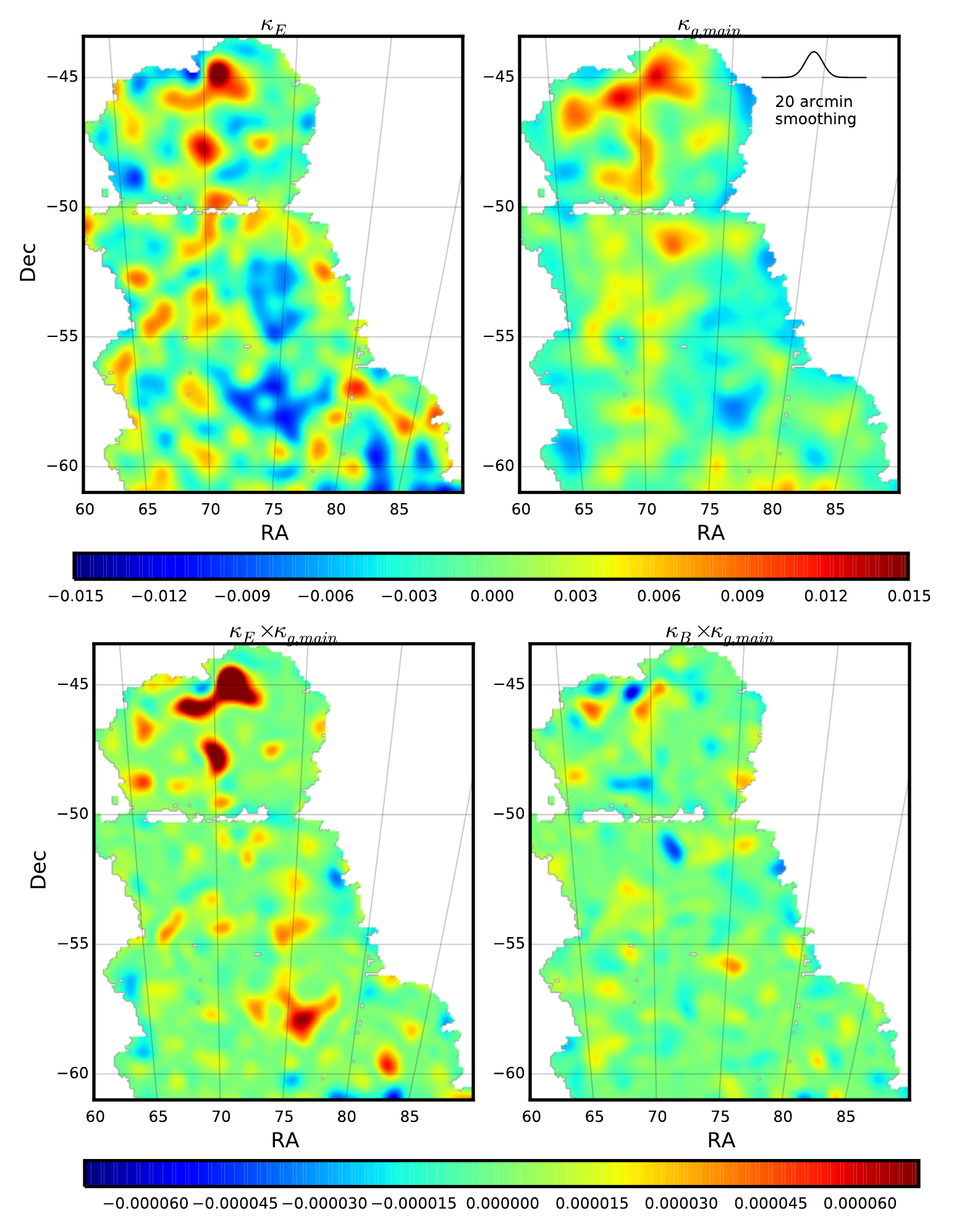}

  \caption{The upper left panel shows the E-modes of the weak lensing convergence map. The upper right shows 
  the weighted foreground galaxy map from the main sample, or $\kappa_{g, main}$. 
  The lower two panels show the product maps of the E-mode (left) and B-mode (right) convergence map with the 
  $\kappa_{g, main}$ map. All maps are generated with a 5 arcmin pixel scale and 20 arcmin Gaussian smoothing.  
  Red areas corresponds to overdensities and blue areas to underdensities in the upper panels. White regions 
  correspond to the survey mask. The scale of the Gaussian smoothing kernel is indicated by the Gaussian profile 
  on the upper right corner.  }
\label{fig:massmap_20_ng}
\end{figure*}

 \begin{figure*}
\includegraphics[scale=1]{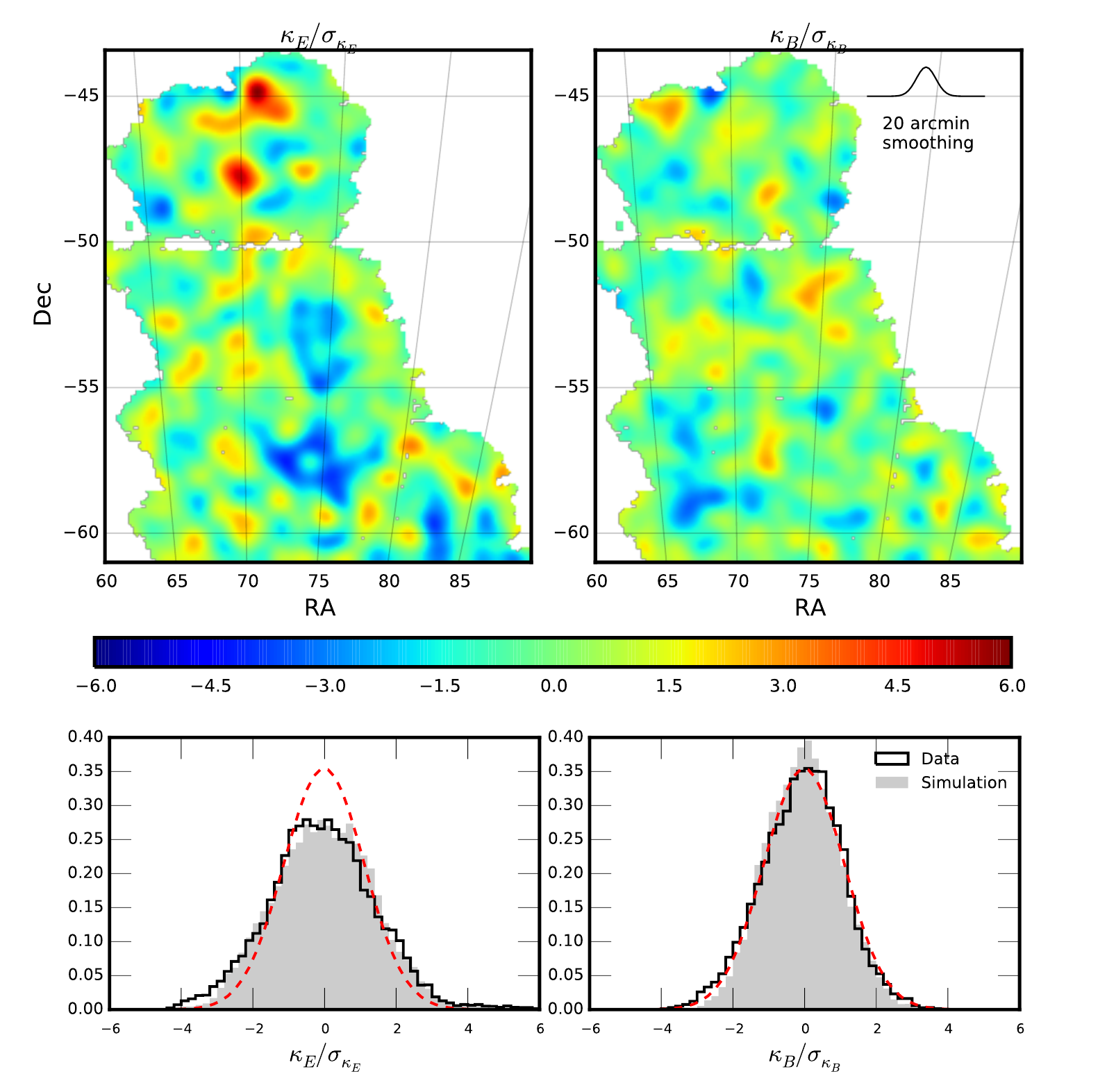}  
  \caption{The top panel shows the S/N map for the mass map in \Fref{fig:massmap_20_ng} estimated via randomized 
  errors. Note that due to the Gaussian smoothing kernel, there is some mixing of 
  scales which leads to higher contrasts in the cores of over and under-dense regions compared to top-hat smoothing. 
  The bottom panel shows the normalized S/N distributions for both maps, overlaid by those measured from simulations 
  described in \Sref{sec:sim_test}. The red dashed lines in both bottom panels show a Gaussian fit to the B-mode S/N.}
\label{fig:s2n_ng}
\end{figure*}

\begin{figure*}
  \begin{center}
  \includegraphics[scale=0.55, angle=-0]{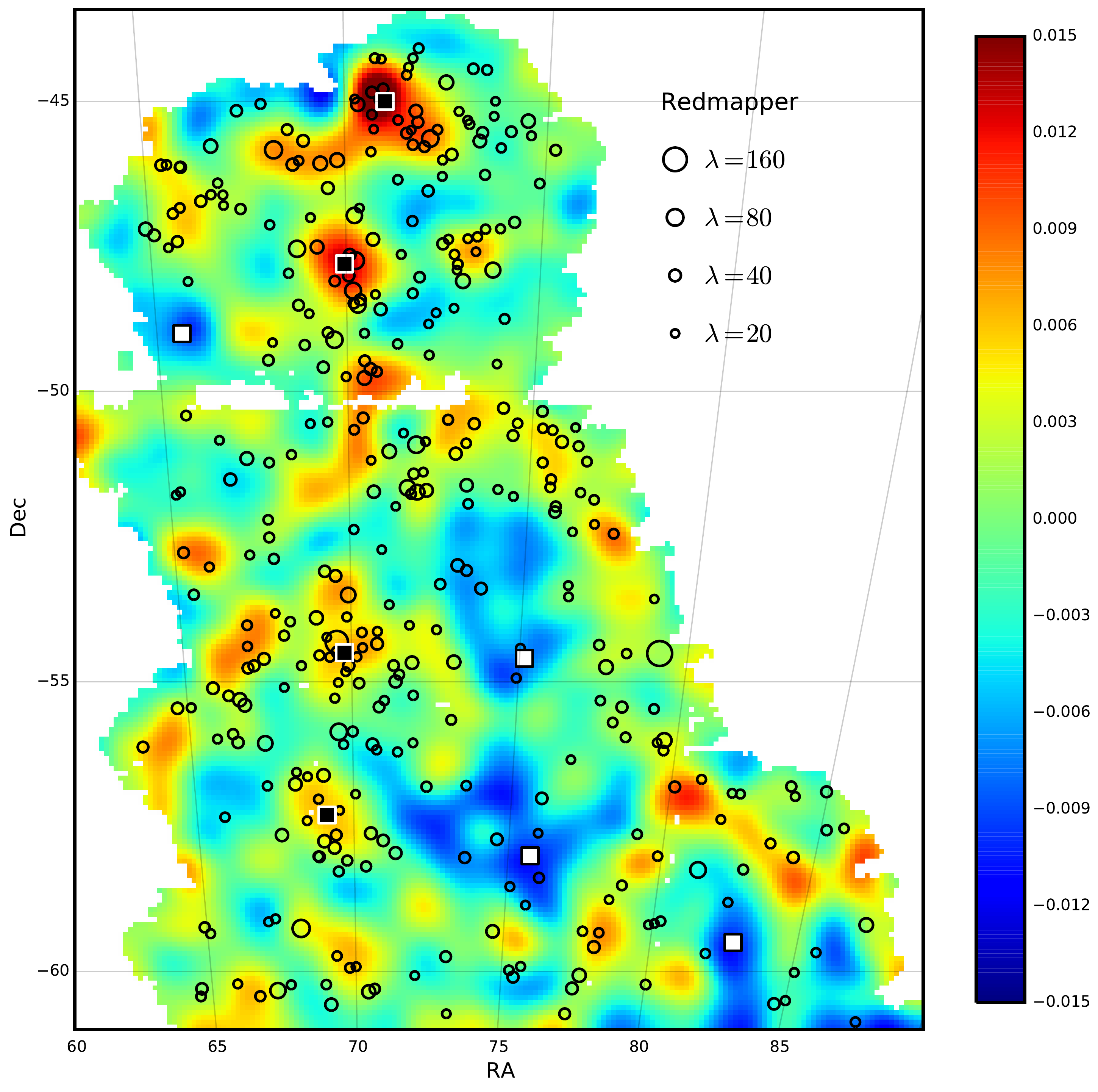}  
    \vspace{-0.3in}
  \end{center}
  \caption{The DES SV mass map along with foreground galaxy clusters detected using the Redmapper algorithm. 
  The clusters are overlaid as black circles with the size of the circles indicating the richness of the cluster. Only 
  clusters with richness greater than 20 and redshift between 0.1 and 0.5 are shown in the figure. The upper right 
  corner shows the correspondence of the optical richness to the size of the circle in the plot.   
  It can be seen that there is significant correlation between the mass map and the distribution of galaxy clusters. 
  Several superclusters (black squares) and voids (white squares) can be identified in the joint map. }
\label{fig:massmap_redmapper_ng}
\end{figure*}

\section{Mass maps}
\label{sec:massmaps}

In \Fref{fig:massmap_20_ng} we show our final convergence maps generated using the data described 
in \Sref{sec:data_sva1} and the methods described in \Sref{sec:ks} and \Sref{sec:fgmaps}. For the purpose of 
visualization we present maps for 20 arcmin Gaussian smoothing. In the top left panel we show the E-mode 
convergence map generated from shear. The top right panel shows the weighted foreground galaxy map 
from the main sample, $\kappa_{g, main}$ map. In both of these panels, red areas correspond to overdensities 
and blue areas correspond to under densities. The bottom left and bottom right panels show the product of 
the $\kappa_{E}$ (left) and $\kappa_{B}$ (right) maps with the $\kappa_{g, main}$. Visually we see that there 
are more positive (correlated) areas for the $\kappa_{E}$ map compared to the $\kappa_{B}$ map, indicating 
clear detection of the weak lensing signal in these maps. Note that these positive regions could be either mass 
over-densities or under-densities. In \Sref{sec:m2l}, we present a quantitative analysis of this correlation.

To estimate the significance of the structures in the mass maps, it is important to understand the noise properties 
of these maps. Uncertainties in the lensing convergence map include contributions from both shape noise and 
measurement uncertainties, which is affected by the number density of galaxies across the field and the shear 
measurement method. 

We estimate the uncertainties on each pixel by randomising the shear measurements on each galaxy. A thousand 
random background galaxy catalogs were generated by shuffling the shear values between all the galaxies. We 
then construct $\kappa_{E}$ and $\kappa_{B}$ maps from these randomized catalogs in the same way as in 
\Fref{fig:massmap_20_ng}. The standard deviation map for these 1000 random samples is used as the noise map. 
Dividing the signal map (\Fref{fig:massmap_20_ng}) by the noise map gives an estimate for 
the S/N of the different structures in the maps, as shown in \Fref{fig:s2n_ng}. These values are broadly consistent 
with those predicted via \Eref{Eq:error} and simulations described in \Sref{sec:sim_test}. The bottom panels of 
\Fref{fig:s2n_ng} show the distribution of the S/N values for both E and B-mode maps for data as well as 
simulations predicted by \Eref{Eq:error}. We find that the B-mode distribution is consistent with a Gaussian 
distribution of standard deviation $\sim1$ as expected \citep{2014ApJ...786...93U}, and the E-mode gives more 
extreme values. The difference between the data and the simulation is consistent with cosmic variance and shape 
noise.

\begin{figure*}
  \begin{center}
 \includegraphics[scale=1]{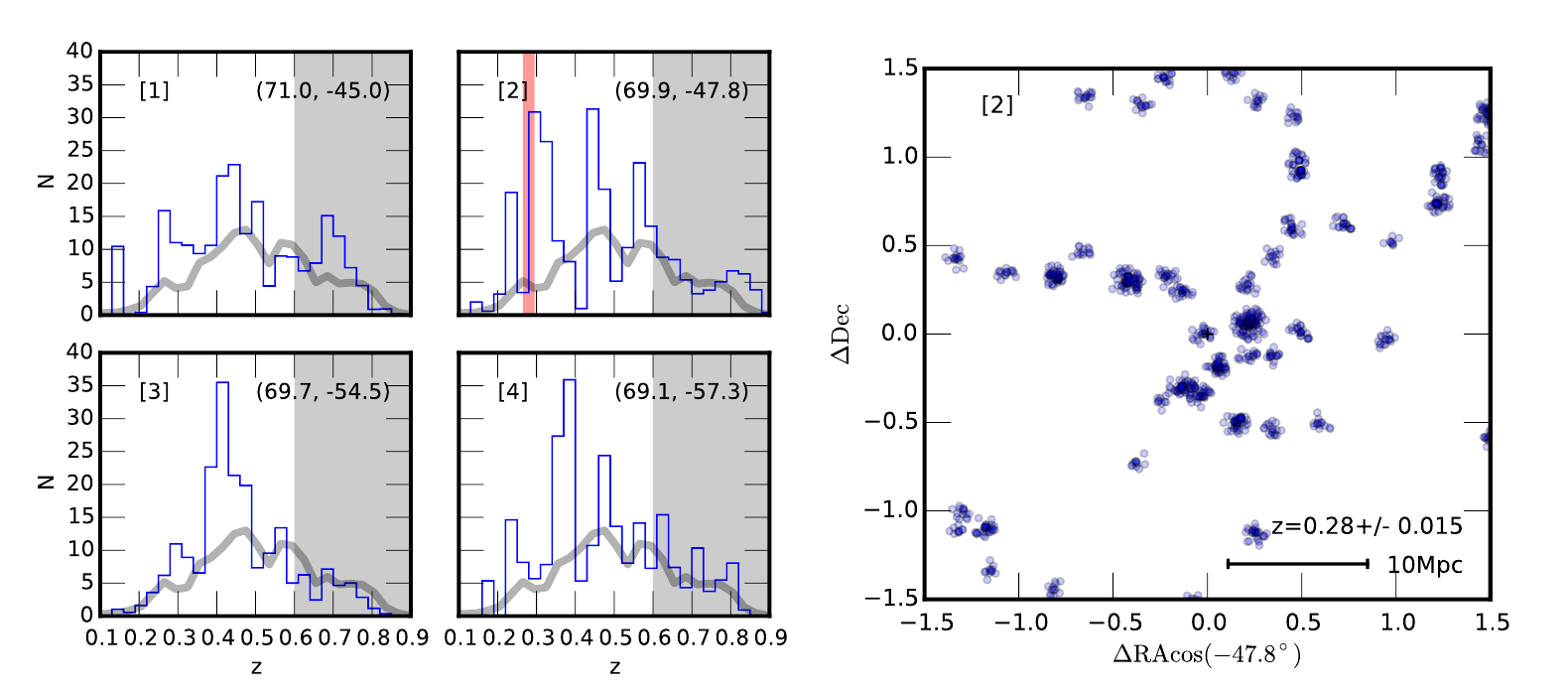}  
      \vspace{-0.2in}
  \end{center}
  \caption{Left: the blue curve in each of the 4 panels shows the weighted redshift 
  distribution of galaxy 
  clusters counts with optical richness $\lambda>$5 at the 4 different locations in the mass map of 
  \Fref{fig:massmap_redmapper_ng} corresponding to large convergence peak locations. The RA, Dec 
  coordinates of these pointings are shown in the top right corner of each panel and the field numbers are listed 
  on the top left corner. The counts are calculated for a 1 deg radius area, and the histograms are weighted by 
  $\lambda$ and the lensing efficiency to properly represent the mass distribution and the lensing probed by the 
  mass map. The thick grey line indicates the corresponding average number count in the full map.  
  The redshift range above $z = 0.6$ is marked with the shaded grey area, as these ranges overlap with the 
  background sample. Right: a candidate supercluster is shown by zooming in on a narrow redshift range of field 2 
  (red band in upper right panel on the left) where a peak in the cluster counts occurs. Each circle indicates the 
  location of a galaxy belonging to a Redmapper cluster. The large spatial extent (a transverse distance of  
  10 Mpc is indicated in the panel) and the irregular shape characteristic of 3D superclusters is evident. 
  }
\label{fig:supercluster_hist}
\end{figure*}

\begin{figure*}
  \begin{center}
 \includegraphics[scale=1]{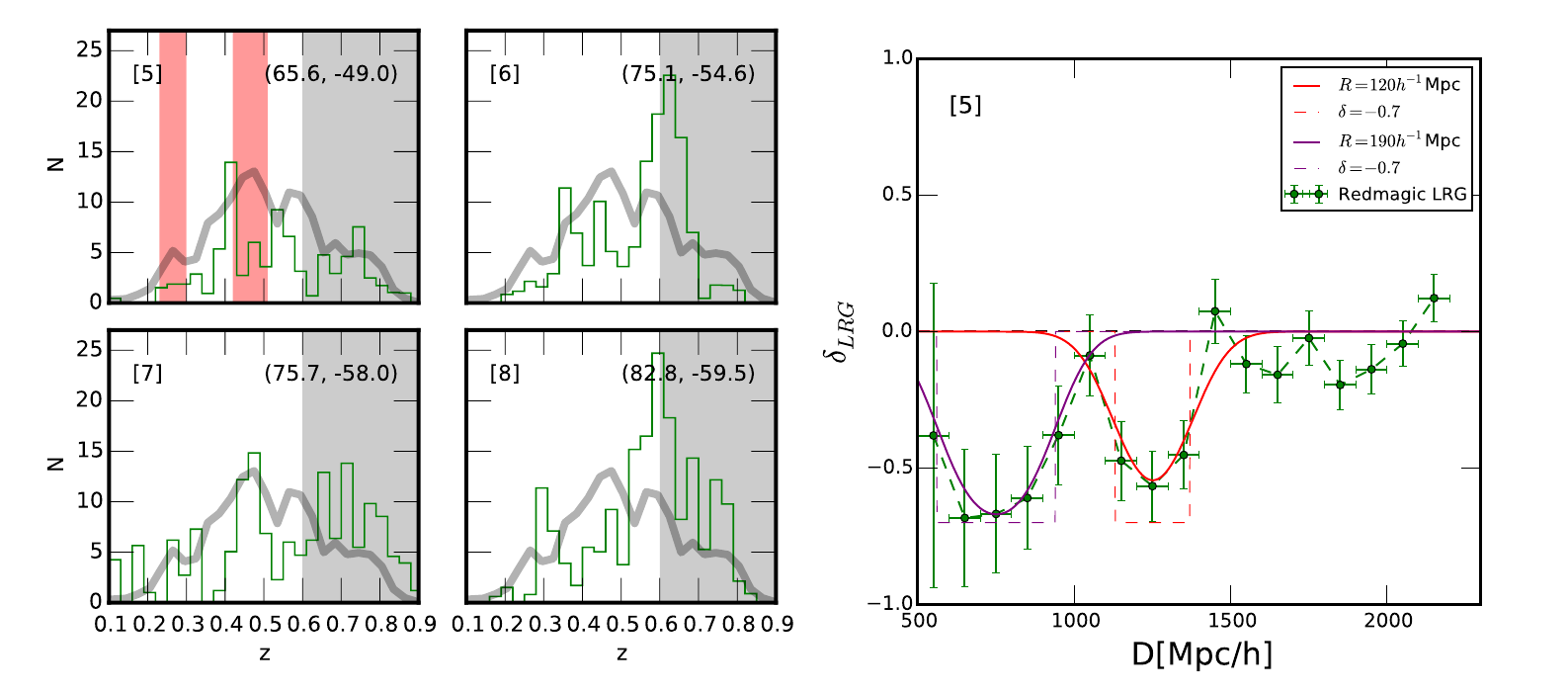}  
      \vspace{-0.2in}
  \end{center}
  \caption{Left: same as the left panel of \Fref{fig:supercluster_hist} but plotted for voids identified in the mass map.
There are typically fewer than average clusters over much of the line of sight which also contains some deep 
underdense regions at specific redshifts. At the higher redshifts, there are also above average cluster counts, 
but since the redshift range overlaps with the source galaxy sample, the interpretation of the structures is 
more complicated. 
 Right: radial distribution of the Redmagic LRGs for field 5 in the left panel (red bands in upper left panel). The data 
 are consistent with the existence of two voids modeled by the ``top-hat'' void model of width 190 Mpc/$h$ and 
 120 Mpc/$h$ respectively.}
\label{fig:supervoid_hist}
\end{figure*}

\section{Correlation with galaxy clusters and potential super-structures}
\label{sec:corr-rm-cluster}

In this section we compare our mass map with optically identified groups and clusters of galaxies using 
the Redmapper algorithm (Rykoff et al. in preparation) on DES data. 
We overlay in \Fref{fig:massmap_redmapper_ng} Redmapper clusters and groups on the 
mass map as black circles. The size of these circles corresponds to the optical richness of these structures. Only 
clusters with optical richness $\lambda$ greater than 20 and redshift between 0.1 and 0.5 are shown in the figure. 
According to \citet{2012ApJ...746..178R} and Saro et al. (in preparation), cluster mass scales approximately linear 
with $\lambda$, with $\lambda=20$ corresponding to $\sim 1.7 \times 10^{14}$ M$_{\odot}$ and $\lambda=80$ 
corresponding to $\sim 7.6 \times 10^{14}$ M$_{\odot}$. 
It is evident from this figure that the structures in the weak lensing mass map have significant correlation with the 
distribution of optically identified Redmapper clusters. 
The combination of the lensing mass maps, Redmapper clusters and Redmagic LRGs provides a powerful 
tool for identifying superstructures in the universe that would otherwise be hard to spot.

Superclusters are the largest distinct structures in the universe, typically 10 Mpc or larger in extent with 
fractional overdensities of order 1-10 times the cosmic mean density. Cosmic voids are the corresponding 
underdensities, typically larger than 10 Mpc in radius with fractional underdensity of order unity.
We identify superclusters and voids from the mass and galaxy maps in \Fref{fig:massmap_20_ng} and 
\Fref{fig:massmap_redmapper_ng}. The large peaks at the positions (RA, Dec) = (71.0, -45.0), (69.9, -47.8), 
(69.7, -54.5) and (69.1, -57.3) and large voids at (RA, Dec) = (65.6, -49.0), (75.1, -54.6), (75.7, -58.0) and 
(82.8, -59.5) are selected as shown in \Fref{fig:massmap_redmapper_ng}. 
The transverse spatial extent of these superstructures is typically greater than 10 Mpc. We compare in the left panels 
of \Fref{fig:supercluster_hist} and \Fref{fig:supervoid_hist} the redshift distribution of the foreground clusters within 1 deg 
radius of these locations with the average redshift distribution of the clusters in the entire SV field. The histograms are weighted 
by the optical richness $\lambda$ as well as the lensing efficiency of our source sample (\Fref{fig:photoZ_dist_sample_ng}). 
$\lambda$ scales roughly linearly with the total mass of the cluster \citep{2012ApJ...746..178R}. 
We find that some of the mass map peaks correspond to supercluster-like structures that are localized in narrow redshift 
ranges, while others (e.g. field 3) show evidence for structures extending over wider redshift range.
On the other hand the large voids 
typically have fewer clusters than average along the line of sight and some deep underdense regions (candidate  
3D voids) at specific redshifts. In some cases there are also above average cluster counts in small ranges 
in redshift (field 6), as expected from the projected nature of these mass maps. The redshift range above $z=0.6$ is marked 
with the shaded grey area, as this range overlaps with the background sample thus complicating the interpretation of the 
relationship with the mass map. In the future we will carry out more detailed studies of the mass maps 
using lensing tomography. 

We show two cases for further investigations of potential superclusters and voids identified through this method. 
First, we look at the spatial distribution of the cluster members in thin redshift slices, identical to the analysis in 
\citet{2015MNRAS.449.2219M}, and find structures such as the one shown in the right panel of \Fref{fig:supercluster_hist}. 
The redshift extent $\Delta z =$0.03 corresponds to a line-of-sight distance of about 90 Mpc/$h$, while the transverse 
size of the structure shown on the right is about 20 Mpc/$h$. 
The line of sight scale  corresponds to the size of the largest 
filamentary structures in cosmological simulations \citep{2014MNRAS.441.2923C}.
These numbers indicate that this is a good candidate for a 3D supercluster. The tight photo-$z$ accuracy of the Redmapper 
clusters ($\sigma_z \approx 0.01(1+z)$) gives us confidence in the identification of real 3D structures. 

For the voids, 
we follow the method developed in \citet{2015MNRAS.450..288S} and study the radial distribution of the foreground Redmagic 
LRGs. We use LRGs within 0.5 deg radius of the chosen position and calculate 
$\delta_{LRG}=(n_{LRG}-\bar{n}_{LRG})/\bar{n}_{LRG}$ in 100 Mpc/$h$ radial bins, where $n_{LRG}$ is the 
number of LRGs in that bin and $\bar{n}_{LRG}$ is the average number of LRGs for the full Redmagic 
catalog in the same radial bin. The radial profile for one void is shown in the right panel of \Fref{fig:supervoid_hist}: it 
is consistent with two large voids in this line of sight. We use a simple ``top-hat'' void model \citep{2015MNRAS.450..288S} 
with an amplitude $\delta_{LRG} = -0.7$, an extent of 190 Mpc/$h$ at a distance of 750 Mpc/$h$ for the first void, 
and another one with $\delta_{LRG} = -0.7$, an extent of 120 Mpc/$h$ at 1250 Mpc/$h$. The combination of these two void 
models, smoothed by the photo-$z$ uncertainty, matches well with the data. We also observe that 
there could be a similarly large but shallower void at higher redshift, also contributing to the projected underdensity 
in the mass map. 

The size and mass of the superclusters are of interest for cosmology as they represent the most massive end of the matter 
distribution. The is especially interesting as the DES dataset allows us to extend our studies to $z\approx 1$. We defer more 
detailed studies of superclusters and voids to follow up work. 

\section{Correlation with galaxy distribution}
\label{sec:m2l}

In this section we quantitatively analyze the extent to which mass follows galaxy density in the data. To do this, 
we cross-correlate the weak lensing mass map with the weighted foreground galaxy density map. The correlation 
is quantified via the Pearson cross-correlation coefficient as described in \Sref{sec:sub-ml}. We cross check the
results using simulations in \Sref{sec:sim_test}.

\subsection{Quantifying the galaxy-mass correlation}
\label{sec:sub-ml}

\begin{figure*}
  \begin{center}
  \includegraphics[scale=0.45]{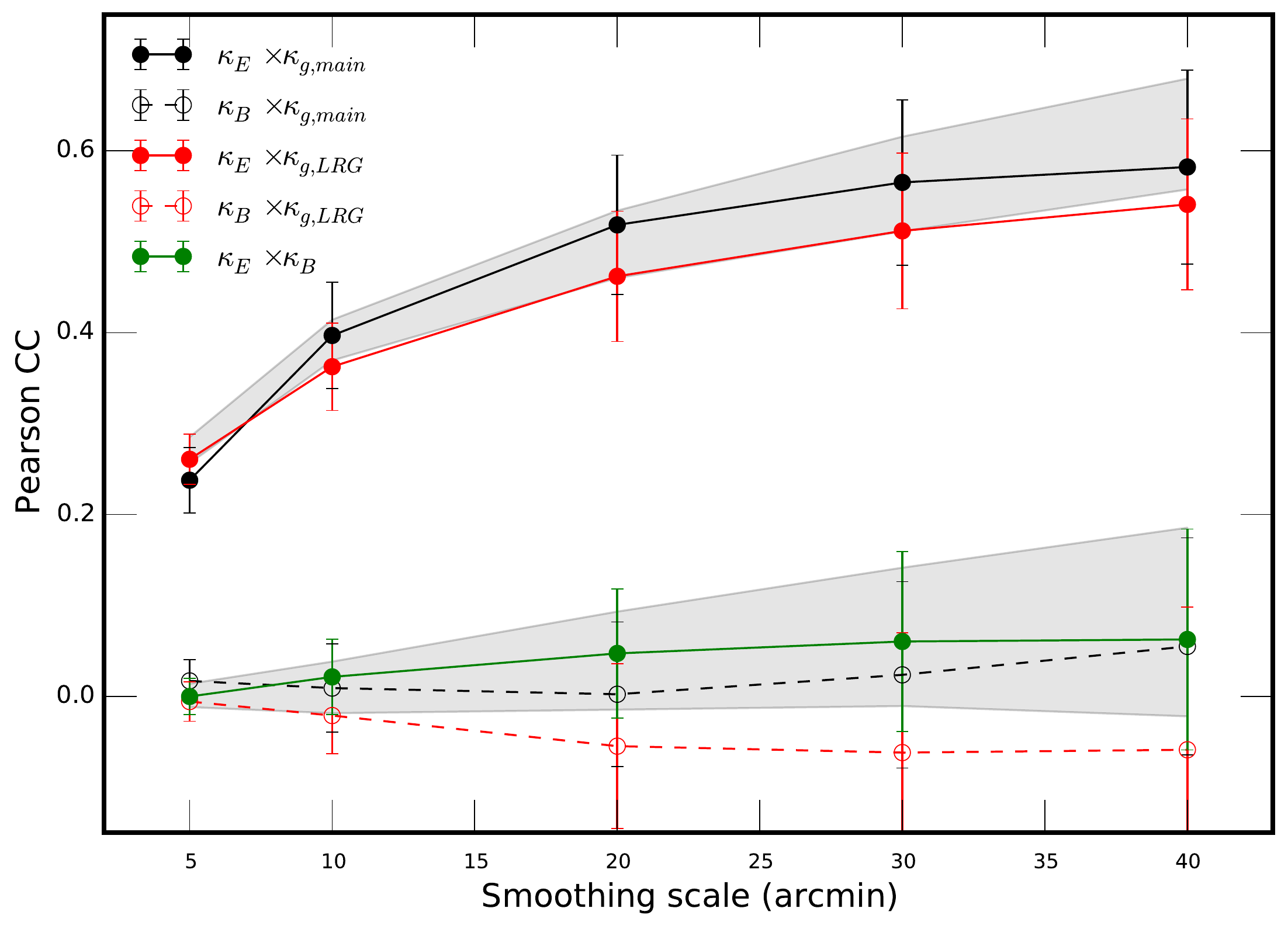}  
  \end{center}
  \caption{This figure shows the Pearson correlation coefficient between foreground galaxies and 
  convergence maps as a function of smoothing scale for the \texttt{ngmix} galaxy catalog. The solid and 
  open symbols show the E 
  and B-mode correlation coefficients respectively. The black circles are for the main foreground sample 
  and the red circles for foreground LRGs. The grey shaded regions show the $1\sigma$ bounds for E 
  and B mode correlations from simulations for the main foreground sample with the same pixelization 
  and smoothing (see \Sref{sec:sim_test} for details). We do not show the similar simulation results for 
  the LRG sample. The detection significance for the correlation is in the range $\sim 5-7 \sigma$ at 
  different smoothing scales. The green points show the correlation between E and B-modes of the 
  mass map. The various B-mode correlations are consistent with zero. Uncertainties on all measurements 
  are estimated based on jackknife resampling. }
\label{fig:cc_vs_scale_ng}
\end{figure*}

\begin{figure*}
  \begin{center}
  \includegraphics[scale=0.45]{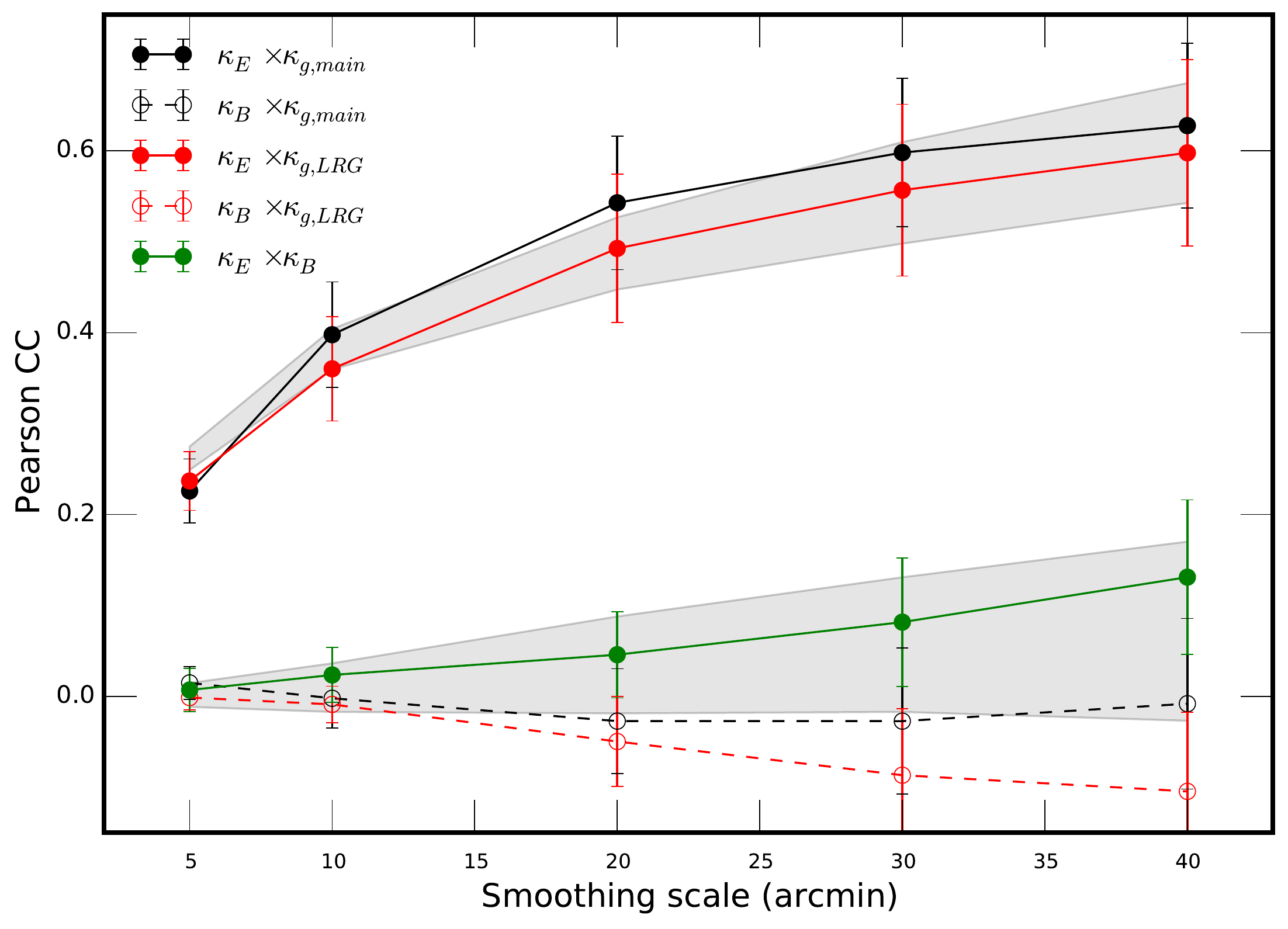}  
  \end{center}
  \caption{Same as \Fref{fig:cc_vs_scale_ng} but using the \texttt{im3shape} galaxy catalog. }
\label{fig:cc_vs_scale_im}
\end{figure*}

We smooth both the convergence maps generated from weak lensing and
from the foreground galaxy density with a Gaussian filter. These
smoothed maps are used to estimate the correlation between the foreground
structure and the weak lensing convergence maps. We calculate the
correlation as a function of the smoothing scale. The correlation is quantified via the Pearson
correlation coefficient defined as
\begin{equation}
\rho_{\kappa_{E} \kappa_g} = \frac{\langle\kappa_{E}  \kappa_g\rangle}{\sigma_{\kappa_{E}} \sigma_{\kappa_g}},
\label{eq:cc}
\end{equation}
where $\langle\kappa_{E}  \kappa_g\rangle$ is the covariance between $\kappa_{E}$ and $\kappa_g$; 
$\sigma_{\kappa_{E}}$ and $\sigma_{\kappa_g}$ are the standard deviation of the $\kappa_{E}$ map, and the $\kappa_g$ 
map from either the foreground main galaxy sample or the foreground LRG sample. In this calculation, pixels in the masked 
region are not used. We also remove pixels within 10 arcmin of the boundaries to avoid significant artefacts from 
the smoothing.

\Fref{fig:cc_vs_scale_ng} shows the Pearson correlation coefficient as function of smoothing scales from 5 to 40 arcmin. 
We find that there is significant correlation between the weak lensing E-mode convergence and convergence from 
different foreground samples, with increasing correlation towards large smoothing scale. This trend is expected for
noise-dominated maps, because the larger smoothing scales reduce the noise fluctuations in the map significantly. A 
similar trend is found when using the LRGs as foreground instead of the general magnitude-limited galaxy sample. 
The lower Pearson correlation between the mass map and LRG sample is because of the larger shot noise due to 
the lower number density compared to the magnitude-limited foreground sample. The error bar on the correlation 
coefficient is estimated based on jackknife resampling. We divide the observed sky into jackknife regions of size 
10 deg$^{2}$ and recalculate the Pearson correlation coefficients, excluding one of the 10 deg$^{2}$ regions 
each time. We found that the estimated uncertainties do not depend significantly on the exact value of patch size. 
We estimate the correlation coefficient after removing one of those patches from the sample to get jackknife 
realizations of the cross-correlation coefficient $\rho_j$. Finally, the variance is estimated as
\begin{equation}
\Delta \rho = \frac{N-1}{N}\sum_j{(\rho_j - \bar{\rho})^2},
\end{equation}
where $j$ runs over all the $N$ jackknife realizations and $\bar{\rho}$ is the average correlation coefficients of all 
patches.
 
We find that the Pearson correlation coefficient between $\kappa_{g}$ from the main foreground galaxy sample 
(LRG sample) and weak lensing E-mode convergence 
is $0.39\pm 0.06$ ($0.36 \pm 0.05$) at 10 arcmin smoothing and 
$0.52 \pm 0.08$ ($0.46 \pm 0.07$) at 20 arcmin smoothing.
This corresponds to a $\sim 6.8 \sigma$ $(7.5 \sigma)$ significance at 10 arcmin smoothing and 
$\sim 6.8 \sigma$ $(6.4 \sigma)$ at 20 arcmin smoothing. As a zeroth-order test of systematics we 
also estimated the correlation between the B-mode weak lensing 
convergence and the $\kappa_{g}$ maps. We find that the correlation between $\kappa_{B}$ and the main foreground 
sample is consistent with zero at all smoothing scales. Similarly, the correlation between E and B 
modes of $\kappa$ is consistent with zero.  
For comparison, we show the same plot calculated for the \texttt{im3shape} catalog in \Fref{fig:cc_vs_scale_im}. We 
find very similar results, with slightly larger correlation between $\kappa_{E}$ and $\kappa_{B}$ at the 1$\sigma$ level. 

\begin{figure}
  \begin{center}
\includegraphics[scale=0.48]{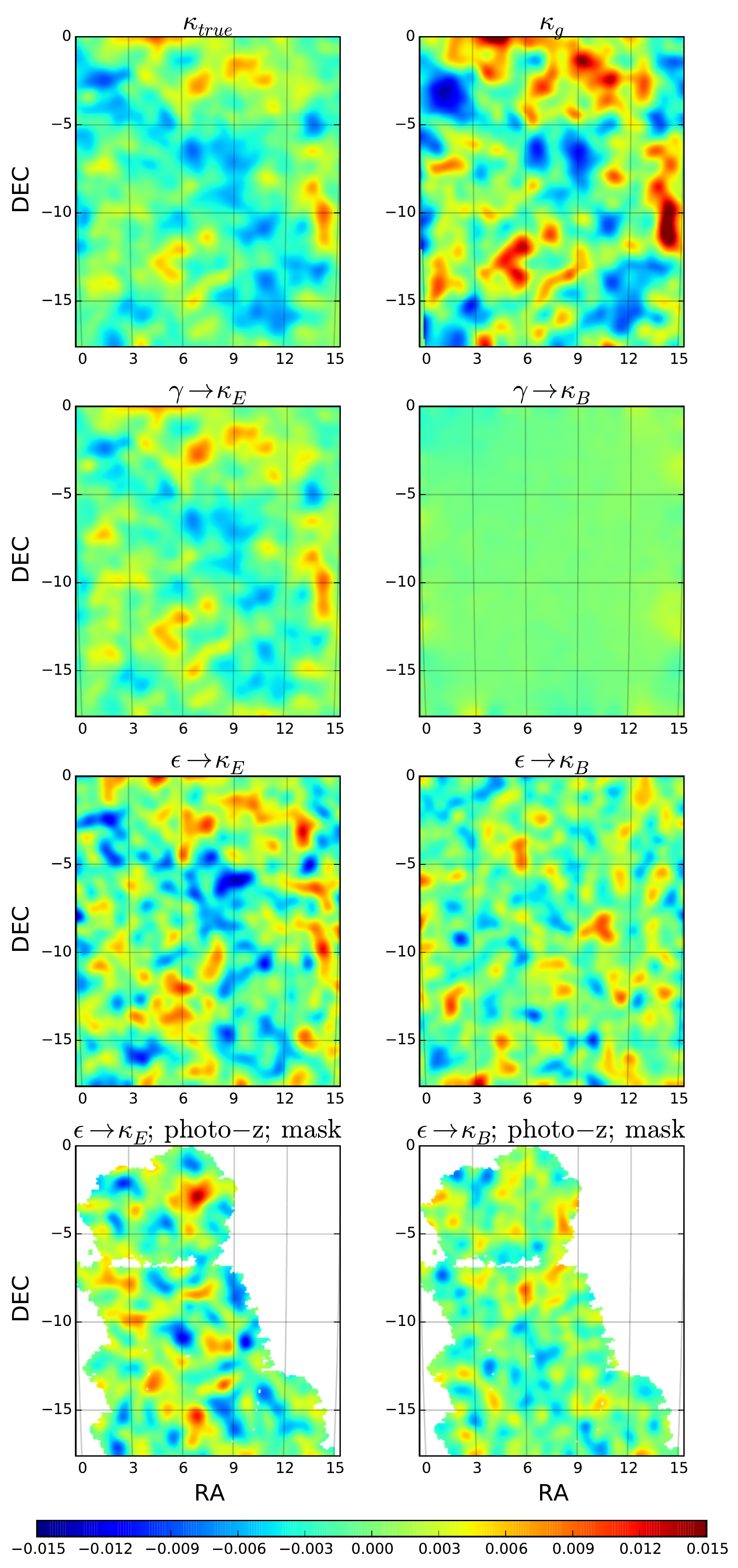}  
   \end{center}
  \caption{Maps from simulations that are designed to mimic the data in our analysis. The simulations are 
  generated for a field of size 15$\times$17.6 deg$^{2}$ with similar redshift and magnitude selections for the foreground 
  and the background sample as the data. The true $\kappa$ and $\kappa_{g}$ maps are shown in the first row, 
  where $\kappa_{g}$ is modelled for the main foreground sample. 
  The reconstructed $\kappa_{E}$ and $\kappa_{B}$ maps from the true 
  $\gamma$ are shown in the first two panels of the second row, followed by the $\kappa_{E}$ and $\kappa_{B}$ 
  maps reconstructed from the ellipticity ($\epsilon$) values. The last row first shows the $\kappa_{E}$ and $\kappa_{B}$ 
  constructed from $\epsilon$ with photo-$z$ uncertainties, then the same maps with an SV survey mask applied. 
  The last two panels on the bottom most closely match the data.}
\label{fig:bcc_sv_ng}
\end{figure}

\begin{figure}
  \begin{center}
  \includegraphics[scale=0.53]{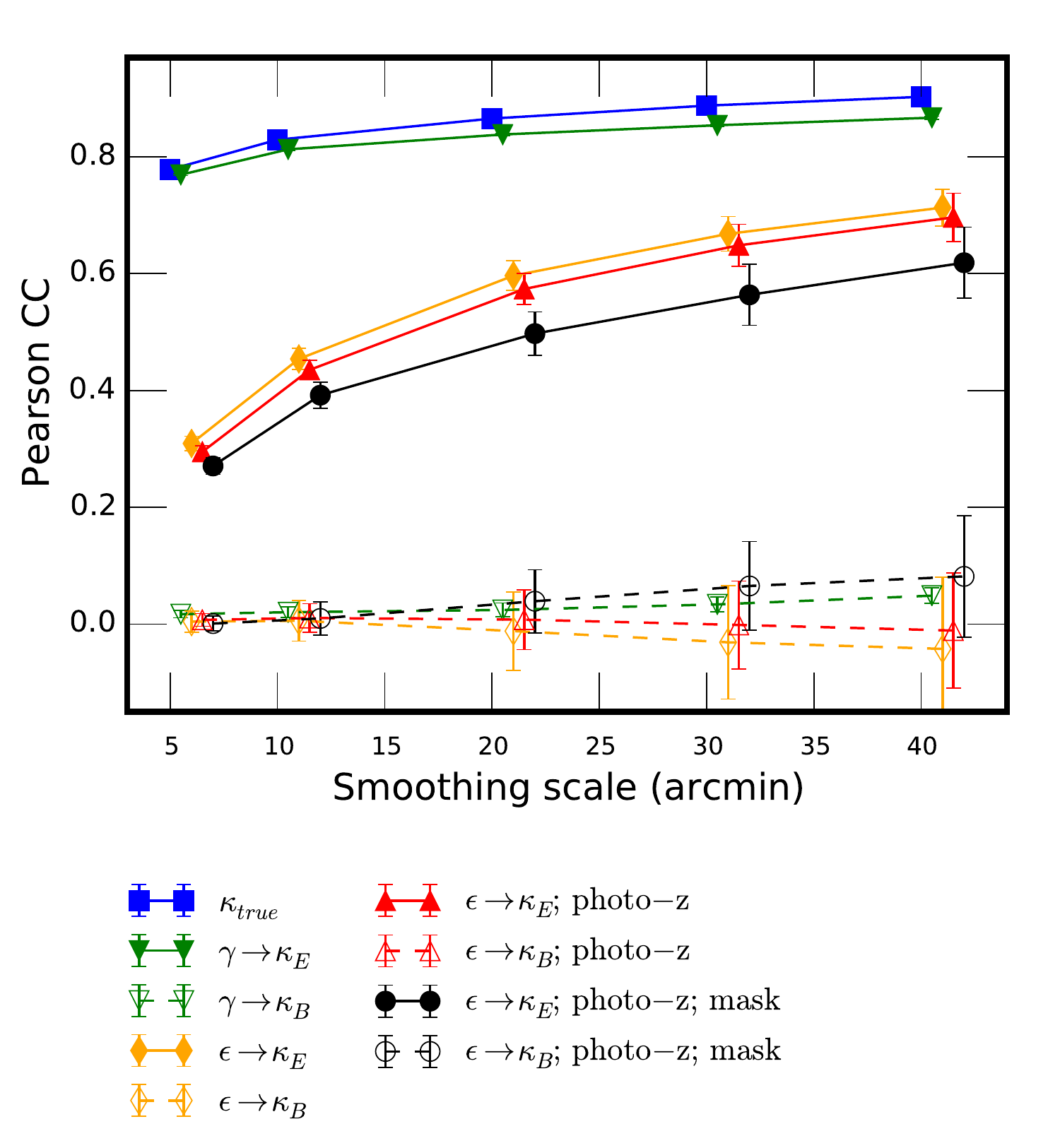}  
  \end{center}
  \caption{Pearson correlation coefficient $\rho_{X \kappa_{g}}$ between the different simulated maps shown 
  in \Fref{fig:bcc_sv_ng} as a function of smoothing scale. $X$ represents the different $\kappa$ maps as listed in the legend. 
  This plot is the simulation version of \Fref{fig:cc_vs_scale_ng}, where one can see how the measured values in the data 
  could have been degraded due to various effects. The qualitative trend of the correlation coefficients as a function 
  of smoothing scale is consistent with that observed in data. When reconstructing $\kappa_{E}$ from the true $\gamma$ 
  small errors are introduced due to the nonlocal reconstruction, lowering the correlation coefficient by a few percent. Adding 
  shape noise to the shear measurement  lowers the signal significantly, with the level of degradation dependent on the smoothing 
  scale. Adding photo-$z$ uncertainties changes the signal by a few percent. Finally, placing an 
  SV-like survey mask changes the signal by $\sim$10\%. The black curve with its error bars corresponds to the shaded region in 
  \Fref{fig:cc_vs_scale_ng}.}
\label{fig:bcc_sv_cc_ng}
\end{figure}

\subsection{Comparison with mock catalogs}
\label{sec:sim_test}

At this point, it is important to verify whether our measurements in the data are consistent with what is 
expected. We investigate this using the simulated catalogs described in \Sref{sec:bcc-data}. As 
the simulations lack several realistic systematic effects in the data, these tests mainly serve as a guidance 
for us to understand: (1) the origin of the B-mode in the $\kappa$ maps, (2) the approximate expected level of 
$\rho_{\kappa \kappa_{g}}$ under pixelization and smoothing, (3) the effect on $\rho_{\kappa \kappa_{g}}$ 
from photo-$z$ uncertainties and cosmic variance, and (4) the effect on the maps and $\rho_{\kappa \kappa_{g}}$
from the survey mask.

We construct a sample similar to the SV data. The same redshift, magnitude, and number density selections in
\Tref{tab:sample_selection} are applied to the simulations to form a foreground and a background sample. 
We choose to simulate the main foreground sample as the LRG foreground sample selection in the 
simulations is less controllable. For the background sample, we add Gaussian noise with standard deviation 
$\sigma=$0.27 to each component of the true shear in the simulations to generate a model for the ellipticities that 
matches the data (Jarvis et al. in preparation).
We then create a $\kappa_{g}$ map from the main foreground sample and a $\kappa$ map from the background 
sample the same way as is done in the data. The cross-correlation coefficient $\rho_{\kappa \kappa_{g}}$ is
calculated from these simulated maps as in \Sref{sec:m2l}. We consider the same range of smoothing scales for 
the maps when calculating $\rho_{\kappa \kappa_{g}}$ as that in \Fref{fig:cc_vs_scale_im}.

The simulations provide us a controlled way of separating the different sources of effects. We construct the 
maps in the following steps, in order of increasing similarities to data: (1) pixelating and smoothing the true 
$\kappa$ values; (2) constructing the $\kappa$ values from the true $\gamma$ values; (3) construct the 
$\kappa$ values from the galaxy ellipticities which include shape noise (we generate 20 realizations); 
(4) using a photo-$z$ model for the foreground and the background galaxies instead of the true redshift; 
generate four different maps from different regions on the sky; (6) use the SV survey mask. 
Note that in step (3) we take the galaxy ellipticity to be the sum of a random component (sampled from a 
Gaussian with standard deviation of 0.27) and the lensing shear, this model is designed to match the data, which 
includes the intrinsic shape noise and other measurement noise associated with e.g. the PSF modelling. 
In step (4) we have modelled the photo-$z$ errors from a spectroscopic sample that ran through the same 
photo-$z$ code, taking the spectroscopic redshift to be the ``true'' redshift. 

The difference between step (1) and step (2) measures the quality of the KS reconstruction method. The 
difference between step (2) and step (3) shows the effect of shape noise and measurement noise. Steps (4), 
(5) and (6) then show the effect of photo-$z$ uncertainties, cosmic variance and masking. 
For each SV-size maps, we generate 20 (shape noise)$\times$4 (cosmic variance)$\times$2 (photo-$z$)
$\times$2 (mask)$=$320 corresponding simulations.
 
\subsubsection{Maps from simulations}
 
\Fref{fig:bcc_sv_ng} shows the various maps generated from one particular patch of the simulations in this 
procedure for 5 arcmin pixels and 20 arcmin smoothing scales (consistent with that in \Fref{fig:massmap_20_ng}). 
The amplitude of  $\kappa_{E}$ and $\kappa_{B}$ both become larger than in the true maps when shape noise is added, 
and the resulting $\kappa_{E}$ map has only slightly higher contrast than the $\kappa_{B}$ map. When photo-$z$ 
uncertainties are included, we see that the peaks and voids in the $\kappa_{E}$ maps visibly move around. 
Applying the mask mainly changes the morphology of the structures in the maps around the edges. 
Comparing the last $\kappa_{E}$ panel in \Fref{fig:bcc_sv_ng} and \Fref{fig:massmap_20_ng}, we see that the amplitude 
and qualitative scales of the variation in the $\kappa_{E}$ maps are similar. On the other hand, if we 
compare the $\kappa_{g}$ maps in the simulations with the $\kappa_{g}$ maps in \Fref{fig:massmap_20_ng}, we find 
some qualitative differences between the simulations and the data. The simulation contains more small scale structure 
and low-$\kappa_{g}$ regions compared to the data. 
We do not investigate this issue further here, as the level of agreement in the simulations and the data is sufficient 
for our purpose.

\subsubsection{Correlation coefficients from simulations}

\Fref{fig:bcc_sv_cc_ng} shows the mean Pearson correlation coefficient between the different maps as a function 
of smoothing scales for the 80 sets of simulated maps (4 different areas in the sky and 20 realisations of shape 
noise each). The error bars indicate the standard deviation of these 80 simulations.

We find $\rho_{\kappa_{true}\kappa_g} \approx 0.8-0.9$.
Several 
factors contribute to this. First, the foreground galaxy sample only includes a finite redshift range, and not all 
galaxies that contribute to the $\kappa_{true}$ map. Second, the presence of a redshift-dependent galaxy bias adds 
further complication to the correlation coefficient. The effect of converting from the true shear 
$\gamma$ to convergence lowers the correlation coefficient by about $3\%$. This is a measure of the error in the KS 
conversion under finite area and resolution of the shear fields. The main degradation of the signal comes when shape 
noise and measurement noise is included. Photo-$z$ uncertainties in both the foreground and the background sample 
changes the correlation coefficient slightly. Finally, the survey mask lowers the correlation coefficient by $\sim10\%$.   

The final correlation coefficient after considering all the effects discussed above is shown by the black curve in 
\Fref{fig:bcc_sv_cc_ng} and overplotted as the shaded region in \Fref{fig:cc_vs_scale_ng}. We find that the 
dependence of $\rho_{\kappa \kappa_{g}}$ on the smoothing scale in the simulation is qualitatively and quantitatively 
very similar to that seen in \Fref{fig:cc_vs_scale_ng}.

\section{Systematic effects}
\label{sec:sys}

In this section we examine the possible systematic uncertainties in our measurement. We focus on the cross correlation 
between our weak lensing mass map $\kappa_{E}$ and the main foreground density map $\kappa_{g, main}$. To simplify 
the notation, we omit the ``main'' in the subscript and use $\kappa_{g}$ to represent the main foreground map in this section. 
We investigate the potential contamination from systematic effects on the cross-correlation 
coefficient $\rho_{\kappa_{E} \kappa_{g}}$
by looking at the spatial correlation of various quantities with 
the $\kappa_{E}$ map and the $\kappa_{g}$ map.

As discussed  in \Aref{sec:foreground_select}, there are several  factors that 
can contaminate the $\delta_{g}$ maps. For example, depth and PSF variations in the observed field 
can introduce artificial clustering in the foreground galaxy density map. Although we use  
magnitude and redshift selections according to the tests in \Aref{sec:foreground_select}, 
one can expect some level of residual effects on the $\kappa_{g}$ maps. 
The $\kappa_{E}$ map is constructed from shear catalogs of the background sample, thus 
systematics in the shear measurement will propagate into the $\kappa_{E}$ map. In Jarvis et al. (in preparation), 
extensive tests of systematics have been carried out on the shear catalog. Therefore here we focus 
on the systematics that are specifically relevant for mass mapping and the correlation coefficient 
\Eref{eq:cc}. 

\begin{table}
\begin{center}
\caption{Quantities examined in our systematics tests. }
\begin{tabular}{ll}
\\
\hline
Map name & Description \\ \hline
\textbf{kE (signal)} & $\kappa_{E}$ from $\gamma_{1}$, $\gamma_{2}$ for background sample\\ 
\textbf{kg (signal)} & $\kappa_{g}$ from main foreground sample \\
kB & $\kappa_{B}$ from $\gamma_{1}$, $\gamma_{2}$ for background sample \\
ns\_f & star number per pixel \\
ng\_b & galaxy number per pixel for background sample \\
snr & signal-to-noise of galaxies in \texttt{im3shape} \\
mask & fraction of area masked in galaxy postage stamp \\
g1 & average $\gamma_{1}$ for background sample\\
g2 & average $\gamma_{2}$ for background sample\\
psf\_e1& average PSF ellipticity \\
psf\_e2 & average PSF ellipticity \\
psf\_T& average PSF size (\texttt{ngmix} only) \\
psf\_fwhm & average PSF size (\texttt{im3shape} only)\\
psf\_kE & $\kappa_{E}$ generated from average PSF ellipticity \\
psf\_kB & $\kappa_{B}$ generated from average PSF ellipticity \\
zp\_b & mean photo-$z$ for background sample\\
zp\_f & mean photo-$z$ for foreground sample\\
ebv & mean extinction\\
skysigma & standard deviation of sky brightness in ADU \\
sky & mean sky brightness in ADU \\
maglim & mean limiting $i$-band AB magnitude\\
exptime & mean exposure time in seconds\\
airmass & mean airmass\\
\hline
\end{tabular}
\label{tab:sys_maps}
\end{center}
\end{table}

\begin{figure*}
  \begin{center}
   \includegraphics[scale=0.98]{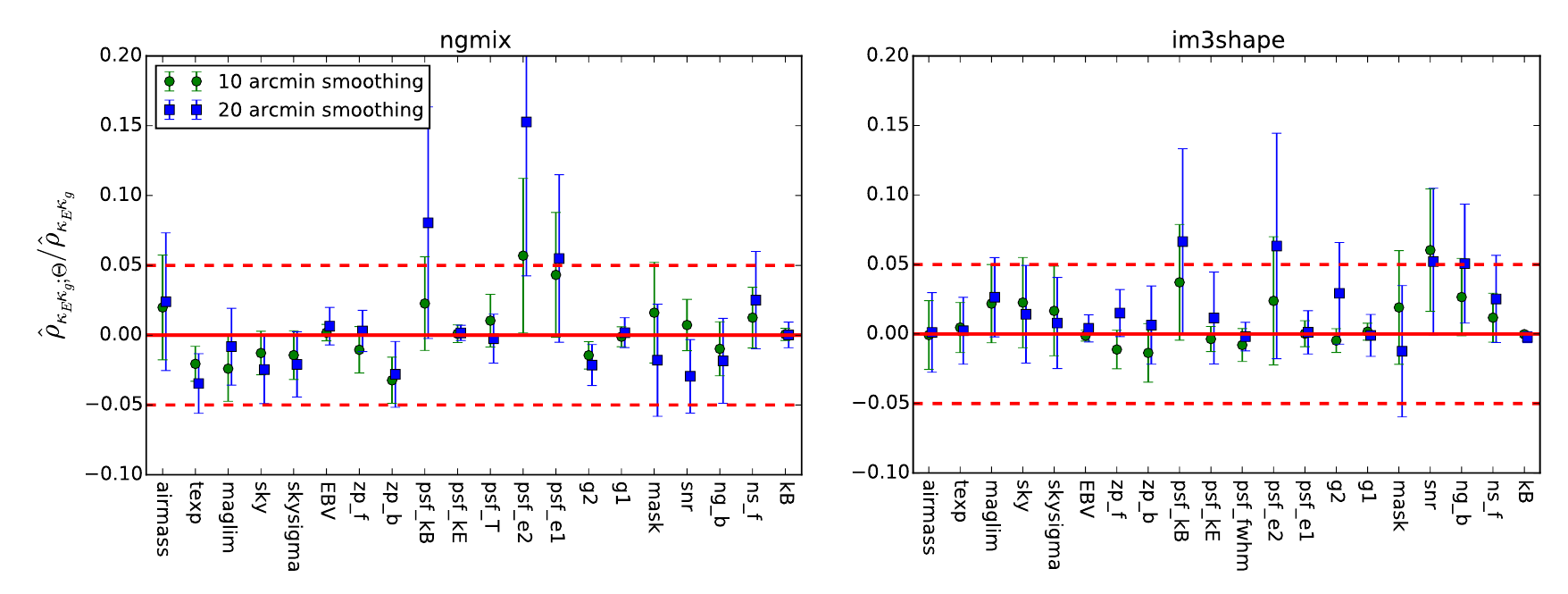}
  \caption{The normalised cross-correlation coefficient $\hat{\rho}_{\kappa_{E} \kappa_{g}; \Theta}$ is shown for 20 
  different systematic uncertainty parameters. The systematics parameters, represented by $\Theta$, are listed 
  in \Tref{tab:sys_maps} and shown for two smoothing scales. The $\hat{\rho}_{\kappa_{E} 
  \kappa_{g}; \Theta}$ values are normalized by $\hat{\rho}_{\kappa_{E} \kappa_{g}}$ to show the relative 
  magnitude of the systematic and the signal. The red dashed line indicates where the systematic is 
  5\% of $\hat{\rho}_{\kappa_{E} \kappa_{g}}$. The error bars are estimated from resampling the foreground and 
  background galaxy sample in patches of size 10 deg$^{2}$. 
  The left panel is calculated for \texttt{ngmix} while the right panel is for \texttt{im3shape}.}
  \label{fig:sys_corr_ng}  
  \end{center}
\end{figure*}

\begin{figure*}
  \begin{center}
  \includegraphics[scale=0.98]{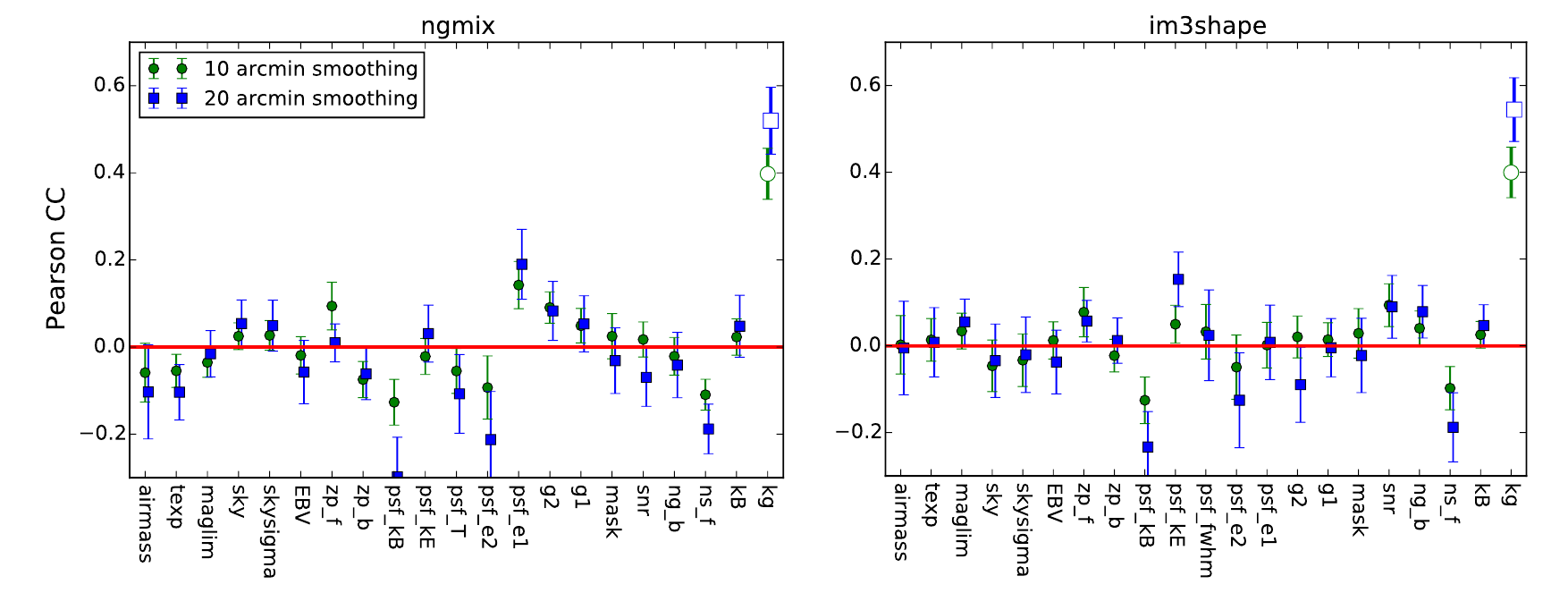}
  \caption{Pearson correlation coefficient $\rho_{\kappa_{E} \Theta}$ where $\Theta$ represents the quantities 
  listed in \Tref{tab:sys_maps}. We show the statistics for two smoothing scales and for both \texttt{ngmix} (left) 
  and \texttt{im3shape} (right). The right-most points in both panel correspond to the detection signal in 
  \Fref{fig:cc_vs_scale_ng} and \Fref{fig:cc_vs_scale_im}. The error bars are estimated from resampling the 
  foreground and background galaxy sample in patches of size 10 deg$^{2}$. Note that this is a different statistic 
  from that in \Fref{fig:sys_corr_ng}, thus the y-axis values are not directly comparable. }
  \label{fig:sys_corr_kE_ng}  
  \end{center}
\end{figure*}

We identify several possible sources of systematics for the background and foreground sample as listed 
in \Tref{tab:sys_maps}. We generate maps of these quantities that are pixelated and smoothed on the 
same scale as the $\kappa_{E}$ and $\kappa_{g}$ maps. 
We then evaluate the contribution of 
these effects to the correlation coefficient (\Eref{eq:cc}) based on the following diagnostic 
quantity:
\begin{equation}
\hat{\rho}_{\kappa_{E} \kappa_{g}; \Theta} = \frac{\hat{\rho}_{\kappa_{E} \Theta} \hat{\rho}_{\kappa_{g} \Theta} }{\hat{\rho}_{\Theta \Theta}}
\label{eq:sys_pearson}
\end{equation}  
with $\hat{\rho}_{XY}$ being the cross-correlation function, which is effectively the unnormalized 
Pearson correlation coefficient between $X$ and $Y$, or
\begin{equation}
\hat{\rho}_{XY} = \langle XY \rangle.
\end{equation}  
Equation \ref{eq:sys_pearson} measures the contribution from some systematics field 
$\Theta$ to $\hat{\rho}_{\kappa_{E} \kappa_{g}}$. 
We calculate $\hat{\rho}_{\kappa_{E} \kappa_{g}; \Theta}$ 
with $\Theta$ being any of the 20 quantities in \Tref{tab:sys_maps} (excluding the signal). \Fref{fig:sys_corr_ng} 
shows the normalized cross-correlation coefficient 
$\hat{\rho}_{\kappa_{E} \kappa_{g}; \Theta}/\hat{\rho}_{\kappa_{E} \kappa_{g}}$ values for all the 
quantities considered for 10 and 20 arcmin smoothing, with the red dashed line at 5\%. The error bars are 
estimated by jackknife resampling similar to that described in \Sref{sec:sub-ml}, and the two panels show 
the results for \texttt{ngmix} and \texttt{im3shape} respectively.
The normalized cross-correlation coefficient is a measure of the fractional contamination in the Pearson 
coefficient (\Eref{eq:cc}) from each of the systematics maps $\Theta$.

We find that for \texttt{ngmix} all quantities show contributions 
to the systematic uncertainties at 10 arcmin smoothing to be at the level of 5\% or lower, while the systematics increase to 
up to 15\% when smoothing at the 20 arcmin scale (though 
with large error bars on the systematics estimation). For \texttt{im3shape}, most of the values stay below 5\% for both 
smoothing scales. The largest contribution in both cases come from the variation in the PSF properties (psf\_e1, 
psf\_e2, psf\_kB). This is expected, as the modelling of the PSF is known to be a significant challenge in weak lensing. 
Since all these PSF quantities are correlated with each other, and many other parameters 
(g1, g2, snr, maglim) are correlated with the PSF properties, we do not expect the total systematics contamination 
to be a direct sum of all these parameters. Instead, we discuss in \Aref{sec:pca_sys} how one can isolate the 
independent contributions of the systematics via a Principal Component Analysis approach and correct for them. 
We find that the correction changes the final Pearson correlation coefficient by 3.5\% relative to the original 
$\rho_{\kappa_{E} \kappa_{g}}$ measured in \Sref{sec:m2l}.

Finally, to check the level of systematic contamination in our $\kappa_{E}$ map itself, we also calculate the 
Pearson correlation coefficient (\Eref{eq:cc}) between the various maps in \Tref{tab:sys_maps} and our 
$\kappa_{E}$ map.  Note that this contamination may or may not be pronounced in \Fref{fig:sys_corr_ng} 
since the statistics plotted there also take into account the correlation of $\kappa_{g}$ with the various quantities. 
This test is independent of the foreground map, therefore is important for applications of the $\kappa_{E}$ map 
that do not also use the foreground maps. \Fref{fig:sys_corr_kE_ng} shows the resulting 21 Pearson correlation 
coefficients. We find that the signal shown in the right-most points in the plot ($\rho_{\kappa_{E}\kappa_{g}}$) is 
larger than all other correlations by at least a factor of $\sim$3.

We also note that in both of these tests, the area of the map is not big enough to ignore the fact that some of these 
correlations can be intrinsically non-zero, even if there were no systematics contamination in the maps.
 
\section{Conclusions}
\label{sec:conclusion}


In this work, we present a weak lensing mass map based on galaxy shape measurements in the 139 
deg$^{2}$ SPT-E field from the Dark Energy Survey Science Verification data. We have cross-correlated the 
mass map with maps of galaxy and cluster samples in the same dataset.  We demonstrate that candidate 
superclusters and voids along the line of sight can be identified exploiting the tight scatter of the cluster 
photo-$z$'s.

We constructed mass maps from the foreground \redmagic\ LRG and general magnitude-limited galaxy 
samples under the assumption that mass traces light. We find that the E-mode of the convergence map
correlates with the galaxy based maps with high statistical significance. We repeated this analysis at various 
smoothing scales and compared the results to measurements from mock catalogs that reproduce 
the galaxy distribution and lensing shape noise properties of the data.  The Pearson cross-correlation coefficient 
is $0.39\pm 0.06$ ($0.36 \pm 0.05$) at 10 arcmin smoothing and $0.52 \pm
0.08$ ($0.46 \pm 0.07$) at 20 arcmin smoothing for the main (LRG)
foreground sample. This corresponds to a $\sim 6.8 \sigma$ $(7.5
\sigma)$ significance at 10 arcmin smoothing and $\sim 6.8 \sigma$
$(6.4 \sigma)$ at 20 arcmin smoothing. We get comparable values from the mock catalogs, indicating that 
statistical uncertainties, not systematics, dominate the noise in the data. The B-mode of the mass 
map is consistent with noise and its correlations with the foreground maps are consistent with zero at the 
1$\sigma$ level.

To examine potential systematic uncertainties in the convergence map we identified 20 possible systematic 
tracers such as seeing, depth, PSF ellipticity and photo-$z$ uncertainties. We show that the systematics 
effects are consistent with zero at the 1 or 2$\sigma$ level. In Appendix B, we present a simple scheme for 
the estimation of systematic uncertainties using Principal Component Analysis. We discuss how these 
contributions can be subtracted from the mass maps if they are found to be significant.

The results from this work open several new directions of study. Potential areas include the study of the 
relative distribution of hot gas with respect to the total mass based on X-ray or SZ observations, estimation of 
galaxy bias, constraining cosmology using peak statistics, and finding filaments in the cosmic web. The tools 
that we have developed in this paper are useful both for identifying potential systematic errors and for 
cosmological applications. The observing seasons for the first two years of DES are now complete 
\citep{2014SPIE.9149E..0VD} and survey an area well over ten times that of the SV data, though shallower 
by about half a magnitude. The full DES survey area will be $\sim 35$ times larger than that presented 
here, at roughly the same depth. The techniques and tools developed in this work will be applied to this new 
survey data, allowing significant expansion of the work presented here.

\section*{Acknowledgements}

We are grateful for the extraordinary contributions of our CTIO colleagues and the DECam 
Construction, Commissioning and Science Verification teams in achieving the excellent 
instrument and telescope conditions that have made this work possible. The success of this 
project also relies critically on the expertise and dedication of the DES Data Management group.

We thank Jake VanderPlas, Andy Connolly, Phil Marshall, Ludo van Waerbeke, and Rafal Szepietowski 
for discussions and collaborative work on mass mapping methodology. 
CC and AA are supported by the Swiss National Science Foundation grants 
200021-149442 and 200021-143906. SB and JZ acknowledge support from a 
European Research Council Starting Grant with number 240672. 
DG was supported by
SFB-Transregio 33 `The Dark Universe' by the Deutsche
Forschungsgemeinschaft (DFG) and the DFG cluster of excellence `Origin
and Structure of the Universe'. 
FS acknowledges financial support provided by CAPES under contract No. 3171-13-2. 
OL acknowledges support from a European Research Council Advanced Grant FP7/291329

Funding for the DES Projects has been provided by the U.S. Department of Energy, the U.S. National Science 
Foundation, the Ministry of Science and Education of Spain, the Science and Technology Facilities Council of 
the United Kingdom, the Higher Education Funding Council for England, the National Center for Supercomputing 
Applications at the University of Illinois at Urbana-Champaign, the Kavli Institute of Cosmological Physics 
at the University of Chicago, the Center for Cosmology and Astro-Particle Physics at the Ohio State University,
the Mitchell Institute for Fundamental Physics and Astronomy at Texas A\&M University, Financiadora de 
Estudos e Projetos, Funda{\c c}{\~a}o Carlos Chagas Filho de Amparo {\`a} Pesquisa do Estado do Rio de 
Janeiro, Conselho Nacional de Desenvolvimento Cient{\'i}fico e Tecnol{\'o}gico and the Minist{\'e}rio da 
Ci{\^e}ncia e Tecnologia, the Deutsche Forschungsgemeinschaft and the Collaborating Institutions in the 
Dark Energy Survey. 

The DES data management system is supported by the National Science Foundation under Grant Number 
AST-1138766. The DES participants from Spanish institutions are partially supported by MINECO under 
grants AYA2012-39559, ESP2013-48274, FPA2013-47986, and Centro de Excelencia Severo Ochoa 
SEV-2012-0234, some of which include ERDF funds from the European Union.

The Collaborating Institutions are Argonne National Laboratory, the University of California at Santa Cruz, 
the University of Cambridge, Centro de Investigaciones Energeticas, Medioambientales y Tecnologicas-Madrid, 
the University of Chicago, University College London, the DES-Brazil Consortium, the Eidgen{\"o}ssische 
Technische Hochschule (ETH) Z{\"u}rich, Fermi National Accelerator Laboratory,
the University of Edinburgh, 
the University of Illinois at Urbana-Champaign, the Institut de Ci\`encies de l'Espai (IEEC/CSIC), 
the Institut de F\'{\i}sica d'Altes Energies, Lawrence Berkeley National Laboratory, the Ludwig-Maximilians 
Universit{\"a}t and the associated Excellence Cluster Universe, the University of Michigan, the National Optical 
Astronomy Observatory, the University of Nottingham, The Ohio State University, the University of Pennsylvania, 
the University of Portsmouth, SLAC National Accelerator Laboratory, Stanford University, the University of 
Sussex, and Texas A\&M University.

This paper has gone through internal review by the DES collaboration.


\appendix
\section{Foreground sample selection}
\label{sec:foreground_select}

As discussed in \Sref{sec:sample}, we consider two factors that can affect the selection of our foreground 
sample -- spatial variation in depth and spatial variation in seeing. If not taken 
care of, these effects will result in apparent spatial variation of the foreground galaxy number density that is 
not due to the cosmological clustering of galaxies. Below we describe tests for each of these and determine 
a set of selection criteria based on the analysis.

\subsection{Depth variation}
\label{sec:depth}
Spatial variation in the depth of the images can cause the apparent galaxy number density to vary, as more or 
less galaxies survive the detection threshold. We would like to construct a foreground galaxy sample which 
minimizes this varying depth effect. A simple solution is to place a magnitude selection slightly shallower than the 
limiting magnitude in all of the areas considered, so that the sample is close to complete in that magnitude 
range. 

We find that in our area of interest, with a magnitude selection at $i<22$, we have 97.5\% of the area that is 
complete to this magnitude limit. We use the 10$\sigma$ galaxy limiting magnitude to define depth, which is a 
conservative measure for the completeness, as we detect many more galaxies below 10$\sigma$. The detail 
methodology of estimating the limiting magnitude of the data is described in Rykoff et al. (in preparation). The 
2.5\% slightly shallower is not expected to yield significant change in our results.  

\subsection{Seeing variation}
\label{sec:seeing}

\begin{figure}
  \begin{center}
   \includegraphics[scale=0.44]{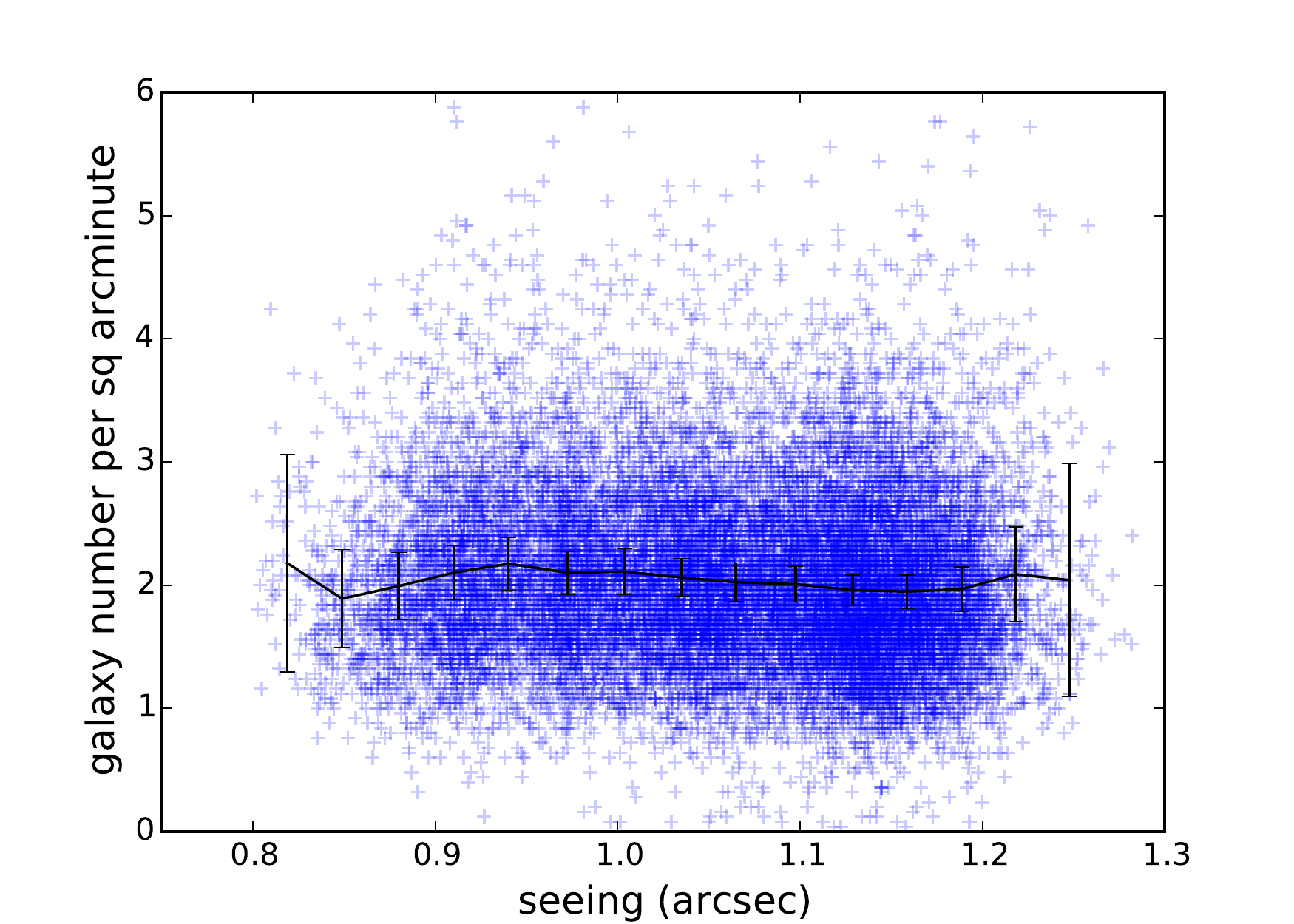}
       \end{center}
  \caption{Galaxy number density as a function of the seeing in the area of consideration. The black line shows 
  the mean and standard deviation (multiplied by 10 for easy visualisation) of the scatter plot in 15 seeing bins. }
\label{fig:seeing}
\end{figure}

Spatial variation in seeing can lead to spatial variation in apparent galaxy number density, as large seeing 
leads to less effective star-galaxy separation as well as higher probability of blending in crowded fields. 
To test this, we first select a foreground sample with $i<22$ and $0.1<z<0.5$ according to \Sref{sec:sample}. 
Then we look at the correlation between the galaxy number density in this foreground sample and 
the average seeing values at these locations, both calculated on a grid of 5$\times$5 arcmin$^{2}$ 
pixels without smoothing. \Fref{fig:seeing} shows the galaxy number density versus seeing. The black data points 
show the mean and standard deviation (multiplied by 10 for easy visualisation) of the scatter plot in 15 seeing bins. There is a 
small anti-correlation between these two values at the 6\% level. This is at an acceptable level for us to continue 
the analysis without masking out the extreme high/low seeing regions. 

\section{Correcting for systematic contamination using PCA}
\label{sec:pca_sys}

As shown in \Sref{sec:sys}, we can use \Eref{eq:sys_pearson} to check for any outstanding 
systematic contamination in our $\kappa_{E}$ map and its correlation with the $\kappa_{g}$ map. Here 
we present a general treatment to correct for these systematic contaminations, similar to that used in 
\citet{2012MNRAS.424..564R} and \citet{2012ApJ...761...14H}.

Assume that our measured $\kappa_{E}$ map is a linear combination of the true $\kappa_{E, {\rm true}}$ map and 
some small coefficient $\alpha_{i}$ times the systematics maps $\{ M_{i} \}$ that can potentially contaminate the 
$\kappa_{E}$ maps (\eg seeing, PSF ellipticity). That is
\begin{equation}
\kappa_{E} = \kappa_{E, {\rm true}} + \sum_{i}^{N} \alpha_{i}M_{i},
\label{eq:sys_kE}
\end{equation}
where we have a total of $N$ systematics maps. 
Similarly, we have the expression for the measured $\kappa_{g}$ in the same way
\begin{equation}
\kappa_{g} = \kappa_{g,{\rm  true}} + \sum_{i}^{N} \beta_{i}M_{i},
\end{equation}
where $\beta_{i}$ is the linear coefficient in this case.

\begin{figure}
  \begin{center}
   \includegraphics[scale=0.48]{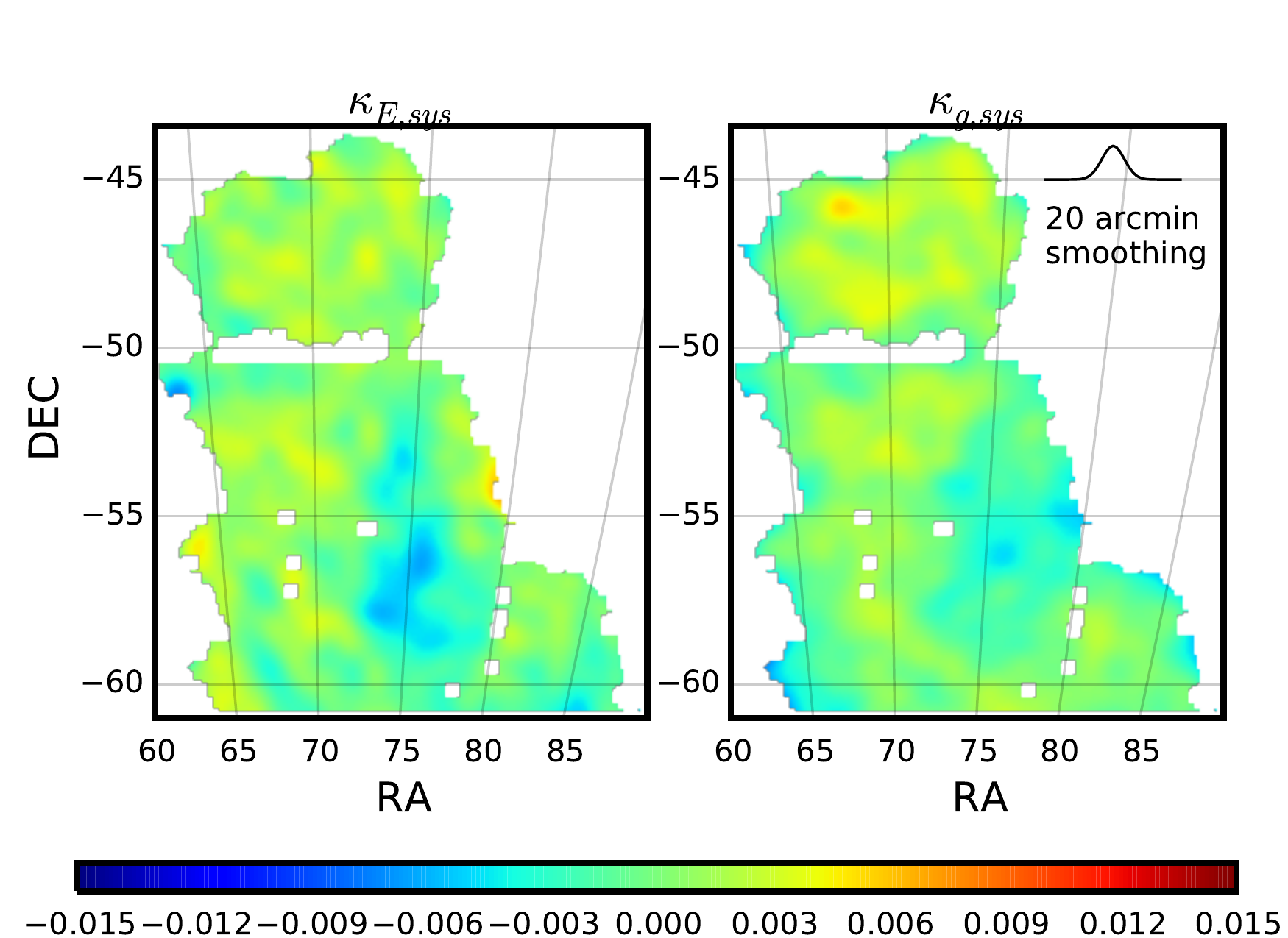}
   \caption{The systematics map for $\kappa_{E}$ (left) and $\kappa_{g}$ (right) shown is compiled using a 
   linear combination of 20 principal components extracted from the systematics maps listed in \Tref{tab:sys_maps}. }
  \label{fig:sys_pca_correct_ng}  
  \end{center}
\end{figure}

Assuming the true maps are uncorrelated with the systematics maps, we have
\begin{equation}
\langle \kappa_{E, {\rm true}} M_{i} \rangle =0;
\end{equation}
\begin{equation}
\langle \kappa_{g, {\rm true}} M_{i} \rangle =0.
\end{equation}
Correlating the measured $\kappa_{E}$ with a single systematics map gives
\begin{equation}
\langle \kappa_{E} M_{j} \rangle= \langle (\sum_{i}^{N} \alpha_{i}M_{i})M_{j}\rangle.
\end{equation}
We can construct a set of systematics maps that are uncorrelated between each other, or
$\langle M_{i} M_{j \neq i} \rangle=0$, and then extract all the coefficients $\alpha_{i}$ from the 
observables as follows:
\begin{equation*}
\langle \kappa_{E} M_{j} \rangle= \alpha_{j} \langle M_{j}M_{j}\rangle;
\end{equation*}
\begin{equation*}
\alpha_{j} = \frac{\langle \kappa_{E} M_{j} \rangle}{ \langle M_{j}M_{j}\rangle};
\end{equation*}
\begin{equation}
\kappa_{E, {\rm true}} = \kappa_{E} - \sum_{i}^{N} \frac{\langle \kappa_{E} M_{i} \rangle}{ \langle M_{i}M_{i}\rangle} M_{i}.
\label{eq:correct_sys_kE}
\end{equation}
And similarly for $\kappa_{g}$, we have
\begin{equation}
\kappa_{g, {\rm true}} = \kappa_{g} - \sum_{i}^{N} \frac{\langle \kappa_{g} M_{i} \rangle}{ \langle M_{i}M_{i}\rangle} M_{i}.
\label{eq:correct_sys_kg}
\end{equation}
To construct a set of systematics maps $\{M_{i}\}$ uncorrelated between each other from a set of 
systematics maps correlated with each other $\{M'_{i}\}$ (\ie\ those listed in \Tref{tab:sys_maps}), 
we invoke the Principal Component Analysis (PCA) method. In this case, each of the pixelated maps, 
after normalizing by its scatter, $\{M'_{i}\}$ form a data vector, and the extracted eigenvectors form a 
orthogonal basis set, which we can use as $\{M_{i}\}$. 
We find that the principal component maps correspond strikingly to physical properties of the data. 
\Fref{fig:sys_pca_correct_ng} shows the systematics maps corresponding to $\kappa_{E}$ and main sample 
$\kappa_{g}$ extracted using this PCA method, or the second terms on the right-hand-side of \Eref{eq:correct_sys_kE} 
and \Eref{eq:correct_sys_kg}. We find that the main contributions come from large-scale 
structures and are at a very low level compared to the original maps (see \Fref{fig:massmap_20_ng}). We 
subtract these systematics maps from the original $\kappa_{E}$ and $\kappa_{g}$ maps according to 
\Eref{eq:correct_sys_kE} and \Eref{eq:correct_sys_kg}. The Pearson correlation coefficient changes 
by 3.5\% relative to the original $\rho_{\kappa_{E} \kappa_{g}}$ measured in \Sref{sec:m2l},
suggesting the contamination to the cross-correlation coefficient is not significant.

It is worth noting that there are a few assumptions that go into the calculation above, which need to be accounted 
for when interpreting these results. First, we have assumed that the systematic maps have no correlation with the 
true $\kappa_{E}$ and $\kappa_{g}$ maps. For a large enough area, this should be true, but for small maps we can 
expect some correlation just by chance. Hence the quantitative ``improvement'' we get in the Pearson correlation 
coefficient must be carefully checked with simulations with larger area than used here. 
Second, since the method is based on PCA, the effectiveness of the correction 
depends on finding the important systematics maps that can contribute \textit{linearly} to the contamination. That is, if 
the systematics come from a non-linear combination of the various maps (\eg\ multiplication of two maps), then one 
would not automatically correct for it without putting in this correct non-linear combination of maps in the first place. 

\end{document}

%